\documentclass{aa}
\usepackage{natbib}
\usepackage{graphicx}
\usepackage{amssymb,times,graphics, subfigure,color,longtable}
\usepackage[english]{babel}
\bibpunct{(}{)}{;}{a}{}{,}

\begin{document}

   \title{Structure of the outer layers of cool standard stars}

   \author{S.~Dehaes \inst{1} \and E.~Bauwens
     \inst{2} \and L.~Decin \inst{2,3}\thanks{\emph{Postdoctoral
         Fellow of the Fund for Scientific Research, Flanders}} \and
     K.~Eriksson \inst{4} \and G.~Raskin \inst{2} \and B.~Butler \inst{5}
     \and C.D.~Dowell \inst{6} \and B.~Ali \inst{7} \and J.A.D.L.~Blommaert \inst{2}}

   \authorrunning{S. Dehaes et al.}

   \offprints{S. Dehaes, \email{sofie.dehaes@wis.kuleuven.be}}

   \institute{
     Department of Mathematics,
     K.U.Leuven, Celestijnenlaan 200B, B-3001 Leuven, Belgium
     \and Department of Physics and Astronomy, Institute for Astronomy,
     K.U.Leuven, Celestijnenlaan 200D, B-3001 Leuven, Belgium
     \and Sterrenkundig Instituut Anton Pannekoek, University of
     Amsterdam, Kruislaan 403 1098 Amsterdam, The Netherlands
     \and Department of Astronomy and Space Physics, Uppsala
     University, Box 515, SE-751\,20 Uppsala, Sweden
   \and NRAO P.O.~Box O, Socorro NM, 87801 USA
   \and Jet Propulsion Laboratory / California Institute of Technology, MS 169-506, 4800 Oak Grove Dr., Pasadena, CA 91109, USA
   \and  IPAC / Caltech, MS 100-22, Pasadena, CA 91125, USA}

   \date{Received; accepted}

   \abstract
{Among late-type red giants, an interesting change occurs
     in the structure of the outer atmospheric layers as one moves to
     later spectral types in the Hertzsprung-Russell diagram: a
     chromosphere is always present, but the coronal emission
     diminishes and a cool massive wind steps in.}
   {Where most studies have focussed on short-wavelength observations,
     this article explores the influence of the chromosphere and the
     wind on long-wavelength photometric measurements. }
 {The
     observational spectral energy distributions are compared with the
     theoretical predictions of the \protect{{\sc marcs}} atmosphere
     models for a sample of nine K- and M-giants. The discrepancies found
     are explained using basic models for flux emission originating
     from a chromosphere or an ionized wind.}
   {For 7 out of $9$ sample stars, a clear flux excess is detected at
     (sub)millimeter and/or centimeter wavelengths, for the other two
     only observational upper limits are obtained. The precise start
     of the excess depends upon the star under consideration. The flux
     at wavelengths shorter than $\sim$1\,mm is most likely dominated
     by an optically thick chromosphere, where an optically thick
     ionized wind is the main flux contributor at longer wavelengths.}
   {Although the optical to mid-infrared spectrum of the studied K-
     and M-giants is well represented by a radiative equilibrium
     atmospheric model, the presence of a chromosphere and/or ionized
     stellar wind at higher altitudes dominates the spectrum in the
     (sub)millimeter and centimeter wavelength ranges. The presence of
     a flux excess also has implications on the role of these stars as
     fiducial spectrophotometric calibrators in the (sub)millimeter
     and centimeter wavelength range. }

   \keywords{ Stars: chromospheres -- Stars: late-type -- Stars: winds, outflows -- Radio continuum: stars   }

   \maketitle


   \section{Introduction} \label{intro}
   Several studies in the past three decades \citep[see, e.g.,][]{linsky1979,
   ayres1981, hunsch1996, haisch1990} have revealed the
   existence of dividing lines in the cool half of the
   Hertzsprung-Russell (HR) diagram, where the giants and supergiants
   reside. These dividing lines are based on differences in the
   physics of the outer atmospheric layers. To the blue side of the
   dividing lines, the late-type stars are surrounded by chromospheres
   and coronae. To the red side the stars also possess chromospheres,
   but in combination with a cool stellar wind. At the time of
   the introduction of the dividing lines \citep{linsky1979,
     ayres1981} there were no observational indications for the
   presence of a corona on the red side, nowadays there is evidence for some
   coronal emission, although much weaker than on the blue  side
   of the dividing lines \citep{ayres1997}.

   Most studies of these outer stellar layers have focussed on X-ray
   and UV observations. But the far-infrared (FIR) continuum can also
   be used as a probe of the layers of the atmosphere for
   solar and cooler stars. As the primary infrared (IR) continuum
   opacity, coming from free-free processes, increases with the square
   of wavelength, we see emission from increasingly outer lying layers
   the longer the wavelength we observe.

   In this study, we use (sub)millimeter and centimeter
   wavelength observations to gain further insight into the outer
   structure of nine giants of spectral type K and M. The selected
   stars are `standard' stars, used in the calibration pedigree of
   many IR spectroscopic and photometric instruments. All of them
   belong to the group of stars with no or low coronal activity.
   With the launch of the ESA-satellite Herschel, which covers
   the full 55 to 672\,$\mu$m wavelength range, it is of interest to
   study whether these low-activity stars can also be used as
   calibrators at these far-infrared wavelengths. In
   Sect.\,\ref{sample} the selection criteria for the stars are
   outlined. Data reduction for the different data-sets is presented
   in Sect.~\ref{datareduction}. Sect.\,\ref{comparison} confronts the
   spectral energy distributions (SEDs) of the stars with the
   hydrostatic {\sc marcs} atmosphere models. In
   Sect.\,\ref{discussion}, the discrepancies found in the previous
   section are discussed. The conclusions are given in
   Sect.\,\ref{conclusions}.


\section{Sample selection}\label{sample}

The stars in our sample are part of a larger set of standard stars
used for the spectrophotometric calibration of near- and mid-infrared
instruments \citep{Decin2003b, Decin2003c, Decin2003d,
  Decin2004ApJS..154..408D, Decin2007A&A...472.1041D,
  Gordon2007PASP..119.1019G}. They are selected for their low chromospheric and coronal activity \citep[see, e.g.,][and further in
Sect.\,\ref{dividing lines} an \ref{wiedemann}]{wiedemann1994, obrien1986}. The position of our sample stars in the HR-diagram indicates that they have low coronal activity, however, it also points to the presence of a cool stellar wind. These stars are potential
candidates to be selected as fiducial calibration sources for the PACS
(60 -- 210\,$\mu$m) and SPIRE (200 -- 670\,$\mu$m) instruments, which
are onboard the ESA Herschel-satellite. Hence, it is necessary to check
if the possible presence of a chromosphere, corona or stellar wind causes
a flux excess in the far-infrared, although they are not visible in the
near-infrared CO lines \citep{wiedemann1994}. Some characteristics of the selected stars are
given in Table~\ref{spectraltype}.

\begin{table}
\begin{center}
\caption{Some characteristics of the sample stars. The values for
  $M_{bol}$ are taken from \citet{Decin2003d}, \citet{kashyap1994} and
  \citet{eggen1970}. No uncertainties were given in these articles on the values of $M_{bol}$ for
   \object{$\iota\,$Aur} and \object{$\sigma\,$Lib}.}
\label{spectraltype}
\begin{tabular}{lccc}
  \hline
  &  spectral type  &    $M_{bol}$       &   $B-V$    \\  \hline\hline
  $\alpha\,$Boo & K2 IIIp   & $-0.90\,\pm\,0.05$ &   $1.23$   \\
  $\alpha\,$Cet & M2 III    & $-3.09\,\pm\,0.13$ &   $1.66$   \\
  $\alpha\,$Tau & K5 III    & $-1.72\,\pm\,0.06$ &   $1.54$   \\
  $\beta\,$And  & M0 III    & $-3.14\,\pm\,0.11$ &   $1.58$   \\
  $\beta\,$Peg  & M2.5 III  & $-3.34\,\pm\,0.11$ &   $1.67$   \\
  $\gamma\,$Dra & K5 III    & $-2.07\,\pm\,0.07$ &   $1.52$   \\
  $\beta\,$UMi  & K4 III    & $-1.71\,\pm\,0.07$ &   $1.51$   \\
  $\iota\,$Aur  & K3 II     & $-2.4$             &   $1.55$   \\
  $\sigma\,$Lib & M3/M4 III & $-3.6$             &   $1.65$   \\
  \hline
\end{tabular}
\end{center}

\end{table}

\section{Observations and data reduction} \label{datareduction}

To construct the SEDs of each of the standard stars several
photometric data points were gathered from the literature. We have
used the UBVRIJKLMNH Photoelectric Catalogue \citep{morel1978}, the
Revised AFGL (RAFGL) Catalogue \citep{price1983}, the IRAS catalogue
of Point Sources (IRAS PSC), Version 2.0 \citep{iras1988},
observations in the Geneva Photometric System 4 \citep{rufener1989},
radio continuum data from \citet{wendker1995} and \citet{cohen2005}, photometric
data in the Johnson's 11-color system \citep{ducati2002}, the COBE
DIRBE Point Source Catalog \citep{smith2004} and the 2MASS All-Sky Catalog of Point Sources
 \citep{skrutskie2006}. A summary of
the available data can be found in Tables~\ref{tablephotometry1} --
\ref{tablephotometry5} in the online appendix.  In order to study the
outer atmospheric layers, (sub)millimeter and centimeter data have
been obtained with \emph{(1.)} the SIMBA bolometer array at $1.2$\,mm
at the SEST telescope, \emph{(2.)}  the MAMBO~II bolometer array at
$1.2$\,mm on the IRAM telescope, \emph{(3.)} the SHARC~II camera at
350\,$\mu$m and 450\,$\mu$m on the CSO, and \emph{(4.)} the VLA at 22 and
43.3\,GHz. The reduction of each of these newly obtained data-sets is
shortly discussed in the next paragraphs.

\subsection{SIMBA observations}

\object{$\alpha\,$Boo}, \object{$\beta\,$And}, \object{$\alpha\,$Cet} and \object{$\beta\,$Peg} were
observed with SIMBA (2003 July 13 -- 15) at $1.2$\,mm, using the fast-scanning
technique. The {\sc mopsi}\footnote{Observers Handbook SIMBA, 2003, edition 1.9,
http://www.ls.eso.org/lasilla/Telescopes/SEST/html/telescope-instruments/simba/index.html}
software developed by R.~Zylka was used for the data reduction. In a first reduction step,
some fundamental operations like despiking, opacity correction and
sky-noise reduction are performed on each scan. Once the scans made
during different nights are assembled, the position of the source is
more accurately determined, which can be used for baseline definition
and for improvement of the sky-noise reduction. For the absolute
calibration, scans of Uranus were used. The model for Uranus is
the standard model offered by {\sc mopsi}, the calibration uncertainties
are estimated at 15\%. After the data reduction, fluxes
were determined using aperture photometry. For each source, the `ideal'
aperture was determined, being the aperture with the highest corresponding
signal-to-noise ratio. A more vast description of the
data reduction and analysis of the SIMBA data is given in
\citet{dehaes2007}. Table\,\ref{simba} shows the determined fluxes together
with the ideal aperture used and the rms noise on the sky background.

\begin{table}
\caption{Fluxes at $1.2\,$mm determined from SIMBA observations. Also listed are the ideal aperture used and the rms noise on the sky background. The given uncertainty does not take the uncertainty on the absolute calibration into account.}
\begin{center}
\begin{tabular}{lccc}
\hline \hline
  target      &  flux               & ideal aperture &   rms    \\
              &  (mJy)              &  (arcsec)      &  (mJy)   \\
\hline
$\alpha\,$Boo &  $105.6\,\pm\,6.9$  & 45             &   $8.4$  \\
$\beta\,$And  &  $<\,40.1$          &                &   $13.8$ \\
$\alpha\,$Cet &  $58.3\,\pm\,4.1$   & 55             &   $6.0$  \\
$\beta\,$Peg  &  $29.2\,\pm\,2.5$   & 20             &   $6.6$  \\
\hline \hline
\end{tabular}
\end{center}
\label{simba}
\end{table}

\subsection{MAMBO~II observations}\label{mambo}
Observations at $1.2$\,mm with MAMBO~II were obtained for \object{$\alpha\,$Boo},
\object{$\iota\,$Aur}, \object{$\beta\,$UMi}, \object{$\gamma\,$Dra}, \object{$\alpha\,$Tau}, \object{$\beta\,$And},
\object{$\alpha\,$Cet}, \object{$\beta\,$Peg} and \object{$\sigma\,$Lib} (2003 October-November).
The reduction was done with an adjusted version of the {\sc mopsi} software called
{\sc  mopsic}\footnote{http://www.iram.es/IRAMES/mainWiki/CookbookMopsic},
using the standard scripts provided for the reduction of On-Off
data. These scripts also include standard reduction steps such as
baseline fitting, despiking, correlated skynoise filtering, etc. Flux
calibration is done using a default conversion factor provided by {\sc
  mopsic}, the calibration uncertainties
are estimated at 15\%. After these reduction steps, scans of the same source are
combined to give one result. Table\,\ref{mambo2} lists the determined fluxes. \\
For $4$ sources both SEST and IRAM data were available. Both measurements coincide for \object{$\beta\,$Peg} and the upper limit determined from the SIMBA observations is in agreement with the flux measured by MAMBO~II for \object{$\beta\,$And}. For \object{$\alpha\,$Boo} and \object{$\alpha\,$Cet}, the 2 measurements do not agree within the errors. As the MAMBO~II observations were performed in service mode and the log-files state very unstable weather conditions for \object{$\alpha$\,Cet} and altocumulus clouds right after the observation of \object{$\alpha$\,Boo}, we have more confidence in the results from the SIMBA observations (which were performed in visitor mode). For \object{$\alpha\,$Boo} and \object{$\alpha\,$Cet} the MAMBO~II observations are discarded for the remainder of the article. Since the MAMBO~II observations for the other objects are in good agreement with the SIMBA data (e.g. \object{$\beta\,$Peg}), with data from the catalogs (e.g. \object{$\beta\,$And}) and since measurements at these long wavelengths are scarce, the data are retained for the other objects.

\begin{table}
\caption{Fluxes at $1.2\,$mm determined from MAMBO II observations. \object{$\alpha\,$Tau} is not listed here, since the uncertainty on the measurement was too large to give any restraints. The observations of \object{$\alpha$\,Boo} and \object{$\alpha$\,Cet} are discarded in the subsequent analysis (see Sect.\,\ref{mambo} for explanation).}
\begin{center}
\begin{tabular}{lccc}
\hline \hline
  target      &  flux (mJy)        \\
\hline
$\alpha\,$Boo &  $20.8\,\pm\,3.5$  \\
$\iota\,$Aur  &  $5.5\,\pm\,1.6$   \\
$\beta\,$UMi  &  $12.2\,\pm\,1.6$   \\
$\gamma\,$Dra &  $10.1\,\pm\,1.3$   \\
$\beta\,$And  &  $23.5\,\pm\,2.7$   \\
$\alpha\,$Cet &  $23.6\,\pm\,2.5$   \\
$\beta\,$Peg  &  $29.5\,\pm\,3.2$   \\
$\sigma\,$Lib &  $12.1\,\pm\,2.0$   \\
\hline \hline
\end{tabular}
\end{center}
\label{mambo2}
\end{table}

\subsection{CSO observations}

Observations of five giant stars were made at $350\,\mu$m and $450\,\mu$m using
the SHARC II camera at Caltech Submillimeter Observatory on several nights
in 2005 and 2008.  Standard
Lissajous scans were used for the stars and calibrators.  The weather
conditions were favorable:  clear skies, low humidity, and precipitable
water vapor in the range $1\,-\,2\,$mm.  Occasional periods of unstable
atmospheric transmission appear to have been properly accounted for in
the data analysis.  Instead of using the facility 225 GHz radiometer for
atmospheric extinction correction, we used (for each observation in
Table~\ref{fluxcso}) the tight correlation between the observed signals from the
calibrators (in raw V) and the average full DC voltage of the bolometers
at the time of the observations to calculate a calibration factor
which was then applied to the target star.  The full DC bolometer voltage
is responsive to the emission from the atmosphere and therefore its
transparency.  In the analysis, the detectors were corrected
for their slight nonlinear response.  The beam size of CSO/SHARC II at 350
and $450\,\mu$m is $8.3\,\pm\,0.3$\,arcsec and $9.8\,\pm\,0.3$\,arcsec, respectively,
and all of the giant stars are unresolved.  The calibrators are unresolved
or only slightly resolved.

The absolute flux calibration is based on the \citet{wright1976} model for
Mars and subsequent planet observations and analysis by \citet{griffin1993}.
  From this work, the absolute uncertainties in the fluxes of Uranus
and Neptune are believed to be $5\,$\%.  Our submullimetre observations of
\object{$\beta$\,}Peg are calibrated directly vs. Uranus and Neptune and are assigned
a systematic calibration uncertainty of $10\,$\% in Table~\ref{fluxcso}.  For the remaining
sources, we used secondary calibrators having fluxes tabulated by \citet{sandell1994},
G. Sandell (priv.\ comm.), \citet{jenness2002}, the
JCMT/SCUBA flux calibration web site (2005 update), and our own cross
calibration work.  Our best estimates for the secondary calibrator fluxes
are given in the table, and the target stars calibrated with respect to them
are assigned a systematic uncertainty of 15\%.  In several cases, the
statistical uncertainties are much smaller than the systematic uncertainties,
so these measurements would benefit from an improved knowledge of the
fluxes of the secondary calibrators.

\begin{table*}
\caption{CSO observations at $350\,\mu$m and $450\,\mu$m.}
\begin{center}
\begin{tabular}{lccccll}
\hline \hline
Star          & Wavelength & Flux  & Statistical & Systematic  & Observing       & Calibrator                    \\
              & ($\mu$m)   & (mJy) & Uncertainty & Uncertainty & Dates           & Fluxes                        \\
              &            &       & (mJy)       & (mJy)       &                 & (Jy/beam)                     \\
\hline
$\alpha$\,Boo & 350        & 601   & 35          &             & 2005 May 10-13  & Arp 220 (10.5)                \\
              &            & 538   & 53          &             & 2008 Mar  1     & Arp 220 (10.5)                \\
              &            & 507   & 19          &             & 2008 May 28     & Arp 220 (10.5)                \\
              &            & 529   & 27          & 79          & average         &                               \\
$\alpha$\,Boo & 450        & 488   & 48          &             & 2008 Mar  2     & Arp 220 (5.4)                 \\
              &            & 440   & 11          &             & 2008 Apr  7     & Arp 220 (5.4)                 \\
              &            & 442   & 11          & 66          & average         &                               \\
$\alpha$\,Cet & 350        & 210   & 16          & 32          & 2008 Sep 22     & Vesta (11.3)                  \\
$\alpha$\,Cet & 450        & 110   & 28          & 17          & 2008 Sep 24     & Vesta (7.1)                   \\
$\alpha$\,Tau & 350        & 530   & 20          & 80          & 2008 Sep 17-18  & Vesta (10.8), Pallas (9.5)    \\
$\alpha$\,Tau & 450        & 304   & 39          & 46          & 2008 Sep 24     & Vesta (7.2), CRL 618 (11.8),  \\
              &            &       &             &             &                 & HL Tau (10.4)                 \\
$\beta$\,Peg  & 350        & 361   &  9          & 36          & 2008 Sep 22     & Uranus (234), Neptune (92),   \\
              &            &       &             &             &                 & CRL 2688 (49)                 \\
$\beta$\,Peg  & 450        & 240   & 12          & 24          & 2008 Sep 23-24  & Uranus (169), CRL 2688 (26.8) \\
$\gamma$\,Dra & 350        & 116   & 25          & 17          & 2008 Sep 22,24  & CRL 2688 (49)                 \\
\hline \hline
\end{tabular}
\end{center}
\label{fluxcso}
\end{table*}

\subsection{VLA observations}

The VLA (Very Large Array) measurements were taken in two bands: the
Q-band (6.9\,cm) and the K-band (1.3\,cm). In all of our
observations, we observed in the continuum mode, which effectively
provides measurements of the total intensity (Stokes I) with an
equivalent bandwidth of $\sim$ 184 MHz~\footnote{The VLA receivers actually operate in the two orthogonal circular
   polarizations, with ~92 MHz bandwidth in each polarization. Since for
   \object{$\alpha$\,Boo} and \object{$\beta$\,Peg}, we expect the two circular polarizations to have
   equal intensity, they are combined into a total intensity
   polarization (Stokes I), for an effective increase of 2 to the
   bandwidth, yielding ~184 MHz equivalent bandwidth. }. The observations for \object{$\alpha$ Boo} were undertaken
on two separate occasions - on January 6, 1999, and on January 25,
2004, for \object{$\beta$\,Peg} on April 21, 2005.  For the 1999 experiment, the
VLA was in the C configuration, with maximum physical antenna
separation of $\sim$3.4\,km.  At this time, only about half of the
antennas were equipped with Q-band receivers and during our experiment
12 were available for this frequency.  The other 15 were tuned to
K-band for simultaneous observations.  The \object{$\alpha\,$Boo} observations were
part of a larger program to observe possible sources for millimeter wavelength flux
calibration that time, and as such were
limited to only about an hour in extent.  For the 2004 and 2005 experiments,
the VLA was in the B configuration, with maximum physical antenna
separation of $\sim$11.4\,km.  A full 6 hour observation was dedicated
to the star at Q-band.

Subsequent calibration of the data proceeded in the normal fashion for
VLA data, in the AIPS reduction package (http://www.cv.nrao.edu/aips/).
For all data, the absolute flux density scale was set with an
observation of 3C286, with assumed flux densities of 1.455 and 2.520\,Jy
for Q- and K-bands, respectively.  Uncertainties in this flux density
scale are $\sim$10\% at Q-band and 5\% at K-band.  Observations of the
unresolved secondary calibrator J1357+193 were used to remove long
timescale (minutes) atmospheric and system fluctuations in the data.
The derived flux densities of J1357+193 were 0.668 and 0.835\,Jy at
Q- and K-bands in 1999, and 1.175\,Jy at Q-band in 2004 (the level
of variation is common with these point-like calibration QSOs at radio
wavelengths).

Images were then constructed from the visibilities via standard AIPS
routines.  The images were lightly CLEANed (a few 10's of components)
to remove the sampling pattern of the array from them.  The final total
flux density was then calculated in five different ways: \emph{(1.)} by counting
up the flux density in the CLEAN components; \emph{(2.)} by taking the peak
flux density in the image; \emph{(3.)} by counting up the flux density around
the central location in the image; \emph{(4.)} by fitting a gaussian to the
image, and taking the peak of that fit gaussian (we do not actually
resolve the star); and \emph{(5.)} by actually fitting the visibilities
themselves to find the flux density of a point source near the image
center.  The final estimated flux density is taken as the median of
these five estimates.  The uncertainty is taken as the average of the
uncertainty from the image and visibility fits.  This is only the
formal uncertainty, systematic uncertainties must be considered in
addition to this.  These can arise from: inaccurate flux density scale,
bad pointing, bad elevation corrections, atmospheric decorrelation,
other electronics sources.  Of these, by far the dominant uncertainty
is the flux density scale, as the others are accounted for in various
ways in the calibration.

Table~\ref{VLAfluxden} shows the resultant flux densities and
uncertainties (formal only) for the VLA observations.  The two
observations of \object{$\alpha$\,Boo} at Q-band are consistent with each other,
and the K-band observation in 1999 is also consistent, given the
expected spectral index.

\begin{table}
\caption{ \label{VLAfluxden}
   Final flux densities from VLA observations of $\alpha$ Boo and
   $\beta$ Peg. The given uncertainty does not take the uncertainty on the absolute calibration into account.
   }
\centering
\footnotesize
\begin {tabular}{ccccc}
\hline\hline
\noalign{\vspace{3pt}}
target & date & frequency & wavelength & flux density \\
       &      &  GHz      & cm         &    mJy       \\
\noalign{\vspace{3pt}}
\hline
\noalign{\vspace{2pt}}
$\alpha$ Boo & 1999-Jan-06 & 22.46 & 6.9 & 1.7$\pm$ 0.2 \\
$\alpha$ Boo & 1999-Jan-06 & 43.30 & 1.3 &3.3 $\pm$ 0.4 \\
$\alpha$ Boo & 2004-Jan-25 & 43.30 & 1.3 &3.34 $\pm$ 0.08 \\
$\beta$ Peg & 2005-Apr-21 & 43.30 & 1.3 &2.49 $\pm$ 0.12 \\
\noalign{\vspace{2pt}}
\hline\hline
\end{tabular}
\end{table}


\section{Comparison between SED and theoretical
  predictions} \label{comparison}

The observational SEDs are compared with the theoretical predictions
of the {\sc sosmarcs} code of \citet{plez1992}, which is a refined
version of the original {\sc marcs} code of
\citet{gustafsson1975}. The synthetic spectra were computed with
\mbox{{\sc turbospectrum}} \citep{plez1992}, the improved version of
the {\sc spectrum} program. For an overview of the continuum and line
opacity lists used, we refer to \citet{decin2000}. The {\sc marcs}
model atmosphere code is built on the assumptions of local
thermodynamic equilibrium, spherical or plane-parallel stratification
in homogeneous stationary layers and hydrostatic equilibrium.

\begin{center}
\begin{figure}
\includegraphics[height=.5\textwidth,angle=90]{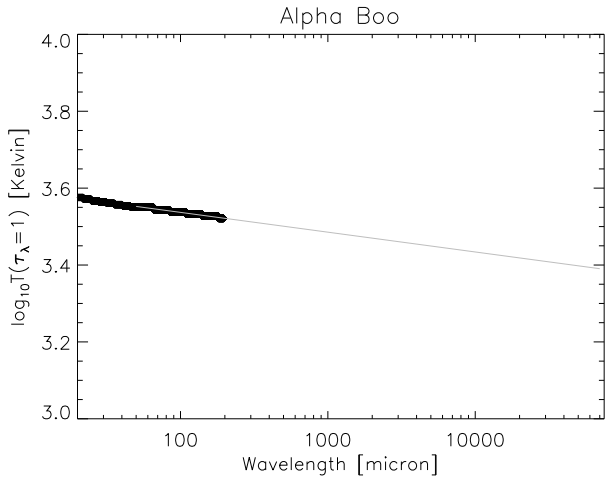}
\caption{Temperature of the atmospheric model layer where
  $\tau_\lambda$\,=\,1 for wavelengths between 20 and 200\,$\mu$m
  (black thick line) for \object{$\alpha$~Boo}. The temperature of
  the characteristic layer where most of the photospheric flux is
  formed ($T_{(\tau_{\lambda}=1)}$) for the full 200 to 7500\,$\mu$m
  wavelength range is derived by extrapolation from the 50 to
  200\,$\mu$m wavelength range (grey line).}
\label{T_extrapol}
\end{figure}
\end{center}

The geometry of the radiation transfer problem for the K- and M-giants
in our sample was given by spherically symmetric layers. Since
the {\sc marcs} model atmosphere only extends up to 200\,$\mu$m, the
far-infrared continuum spectrum was computed by extrapolation from the
continuum theoretical spectrum between 50 and 200\,$\mu$m. We
therefore have determined the temperature of the flux forming region
where $\tau_\lambda = 1$, with $\lambda$ ranging from 50 to
200\,$\mu$m. With H$^-$ free-free being the main continuum opacity
source, subsequent outer cooler layers are sampled for longer
wavelengths. Using a logarithmic extrapolation, the temperature
for the characteristic layer where most of the photospheric flux is
formed ($T_{(\tau_{\lambda}=1)}$) for the full 200 to 7500\,$\mu$m
wavelength range is determined (see Fig.~\ref{T_extrapol}). The
continuum flux at each far-infrared wavelength point is then
approximated by the blackbody flux at the characteristic temperature
$B_\lambda(T)$ scaled with the appropriate angular diameter. Since we
are in the Rayleigh-Jeans part of the spectrum, the flux value is
quite insensitive to the temperature, i.e. $\partial B_\lambda(T)
/ \partial T$ is small.

The accuracy and resolution of today's FIR instruments remain
currently too poor to constrain the importance of line veiling in the
(sub)millimeter range. The study by \citet{Decin2007A&A...472.1041D}
and B.~Plez in case of the 40 -- 665\,$\mu$m spectrum for \object{$\alpha$ Tau}
(\emph{priv. comm.}) shows that molecular line absorption at a
resolution of $\sim$1500 is typically less than 1\,\% beyond
150\,$\mu$m. We therefore will compare the (sub)millimeter
observational data with \emph{continuum} flux predictions.

The input parameters for the {\sc marcs}
code were taken from \citet{Decin2003d} unless indicated otherwise in
Table \ref{inputmarcs}. In the same article a discussion about the
uncertainties on these parameters can be found. The models were
reddened according to the value of the interstellar extinction derived
from the model of \citet{arenou1992} using the distances
from \citet{Decin2003d} or \citet{Ochsenbein1999}. The values of the
interstellar extinction and the distances are listed in Table
\ref{inputmarcs}.

\begin{figure*}
\sidecaption
\includegraphics[width=12cm]{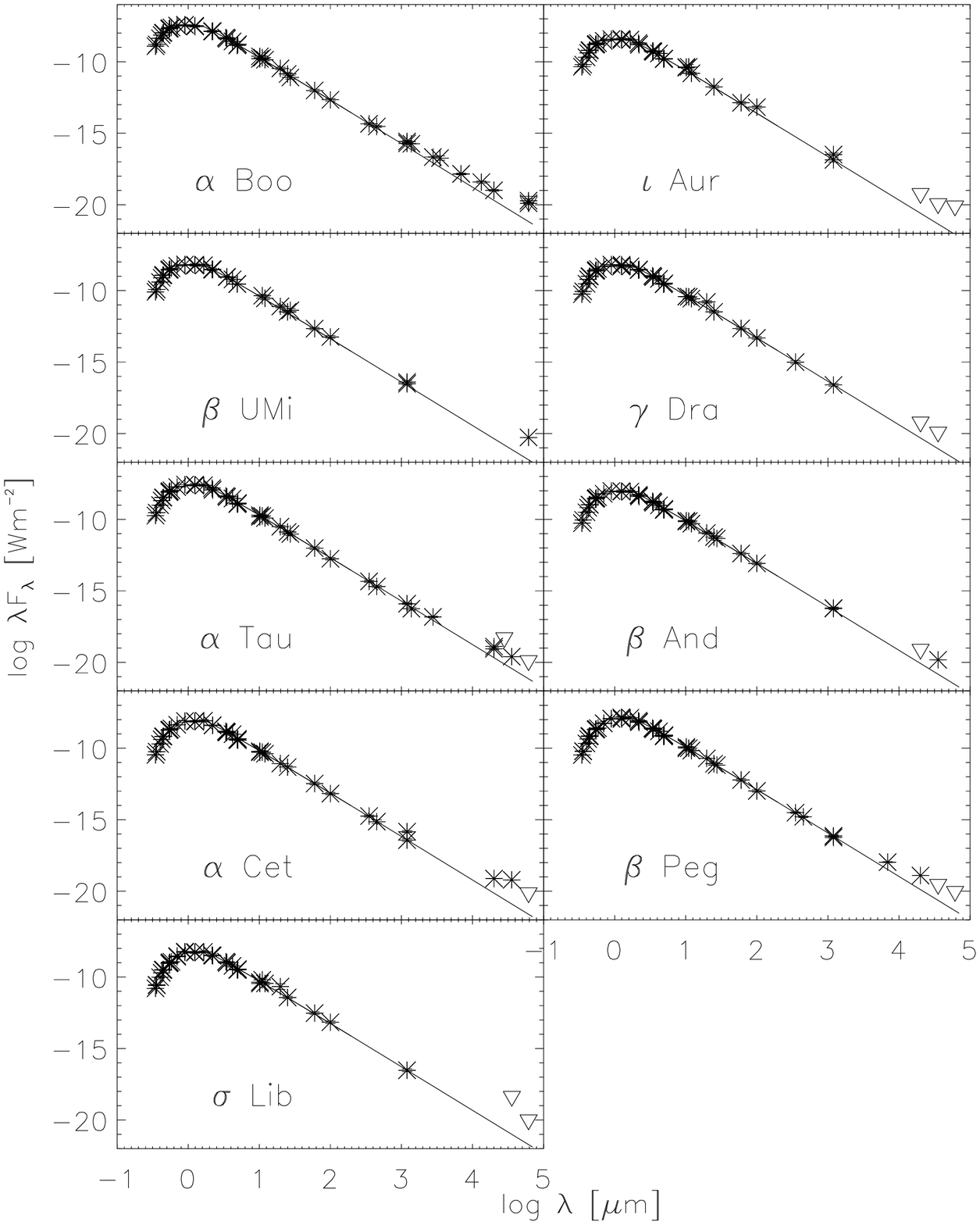}
\caption{Comparison between the photometric data (asterisks) and the
  continuum theoretical {\sc marcs} spectrum (full line) for the nine
  sample stars. If several observations are available at the same
  wavelength, only the maximum and minimum flux value were plotted,
  except at (sub)millimeter and centimeter wavelengths as this
  wavelength region is of particular interest here. Most of the error
  bars fall within the symbols for the data. A reversed triangle
  represents an upper limit.}
\label{sed}
\end{figure*}

Fig.\,\ref{sed} shows the photometric data in comparison with the {\sc
  marcs} models. For all targets, the theoretical predictions
underestimate the observations in the millimeter and/or centimeter
wavelength area. Where an excess is detectable at $1.2$\,mm, the model
underestimates the observations by an average of $25\,\%$. At centimeter
wavelengths, the discrepancy amounts to an average of $90\,\%$. In the
following section, different causes for this excess are explored.

\renewcommand{\arraystretch}{1.1}
\begin{table*}
  \caption{Input parameters for the \protect{{\sc marcs}} code
    from \protect\citet{Decin2003d} unless indicated otherwise:
    the effective temperature $T_{\mathrm{eff}}$ in K, the gravity log
    $g$ in cm/s$^2$, the microturbulent velocity $\xi_{t}$ in
    km\,$\mathrm{s}^{-1}$, the metallicity [Fe/H], the abundances of
    carbon, nitrogen, and oxygen, the $^{12}$C/$^{13}$C-ratio and the photospheric stellar angular diameter $\theta_d$ in
    milliarcseconds. The calculation of the angular diameter is discussed in Sect.\,\ref{proofexcess}. The table  also contains the distances (in pc) and the values of the interstellar
    extinction $\mathrm{A}_\mathrm{v}$ as derived from the model of
    \protect{\citet{arenou1992}}. Values, for which no literature
    values have been found, have been assumed on the basis of analogue
    objects, and are listed in italics.}
\label{inputmarcs}
\centering
\begin{minipage}{\textwidth}
 \centering
\begin{tabular}{l c c c c c c}
\hline
                        & $\alpha$\,Boo       & $\iota$ Aur                          & $\beta$ UMi                        & $\gamma$\,Dra                        & $\alpha$\,Tau      \\
\hline\hline

Sp. Type                & K2 IIIp             & K3 II                                & K4 III                             & K5 III                               & K5 III             \\
$T_{\mathrm{eff}}$      & $4320\,\pm\,140$    & $4160\,\pm\,130$\footnotemark[1]     & $4085\,\pm\,140$                   & $3960\,\pm\,140$                     & $3850\,\pm\,140$   \\
log $g$                 & $1.5\,\pm\,0.15$    & $1.74\,\pm\,0.36$\footnotemark[1]   & $1.6\,\pm\,0.02$                   & $1.30\,\pm\,0.25$                    & $1.50\,\pm\,0.15$  \\
$\xi_{\mathrm{t}}$      & $1.7\,\pm\,0.5$     & $3.00\,\pm\,0.5 $\footnotemark[1]   & $2\,\pm\,0.5 $                     & $2.0\,\pm\,0.5$                      & $1.7\,\pm\,0.5$    \\
${\rm{\left[Fe/H\right]}}$    & $-0.50\,\pm\,0.20$  & $-0.11\,\pm\,0.22$\footnotemark[1]   & $-0.15\,\pm\,0.2$                  & $0.00\,\pm\,0.20$                    & $-0.15\,\pm\,0.20$ \\
$\epsilon$(C)          & $7.96\,\pm\,0.20$   &        \emph{8.35}                              & $8.25\,\pm\,0.2 $                  & $8.15\,\pm\,0.25$                    & $8.35\,\pm\,0.20$  \\
$\epsilon$(N)          & $7.61\,\pm\,0.25$   &        \emph{8.35}                                & $8.16\,\pm\,0.25$                  & $8.26\,\pm\,0.25$                    & $8.35\,\pm\,0.25$  \\
$\epsilon$(O)          & $8.68\,\pm\,0.20$   &        \emph{8.93}                                & $8.83\,\pm\,0.2$                   & $8.93\,\pm\,0.20$                    & $8.93\,\pm\,0.20$  \\
$^{12}C/ ^{13}C$        & $7\,\pm\,2$         &        \emph{10}                              & $9\,\pm\,2$                        & $10\,\pm\,2$                         & $10\,\pm\,2$       \\
$\theta_\mathrm{d}$     & $20.74\,\pm\,0.10$  & $7.05\,\pm\,0.03$                    & $9.03\,\pm\,0.42 $                 & $9.94\,\pm\,0.05$                    & $20.89\,\pm\,0.10$ \\
distance                & $11.26\,\pm\,0.09$  & $166.56\,\pm\,33.31$\footnotemark[5] & $39.87\,\pm\,7.97$\footnotemark[5] & $45.25\,\pm\,0.94$                   & $19.96\,\pm\,0.38$ \\
$M_{g}$                 & $0.73\,\pm\,0.27$   & $3.6$                                & $2.49\,\pm\,0.92$                  & $1.72\,\pm\,1.02$                    & $2.30\,\pm\,0.85$  \\
$\mathrm{A}_\mathrm{v}$ & $0.01\,\pm\,0.15$   & $0.00 \,\pm\,0.15 $                  & $0.00\,\pm\,0.5$                   & $0.03\,\pm\,0.15$                    & $0.03\,\pm\,0.15$  \\

\hline

                         & $\beta$\,And       & $\alpha$\,Cet                        & $\beta$\,Peg                       & $\sigma$ Lib                         &                    \\

\hline\hline
Sp. Type                 & M0 III             & M2 III                               & M2.5 III                           & M3/M4 III                            &                    \\
$T_{\mathrm{eff}}$       & $3880\,\pm\,140$   & $3740\,\pm\,140$                     & $3600\,\pm\,300$                   & $3634\,\pm\,110$\footnotemark[2]     &                    \\
log $g$                  & $0.95\,\pm\,0.25$  & $0.95\,\pm\,0.25$                    & $0.65\,\pm\,0.40$                  & $0.9\,\pm\,0.31$ \footnotemark[2]    &                    \\
$\xi_{\mathrm{t}}$       & $2.0\,\pm\,0.5$    & $2.3\,\pm\,0.5$                      & $2.0\,\pm\,0.3$                    & $3.1\,\pm\,0.5 $\footnotemark[3]     &                    \\
${\rm{\left[Fe/H\right]}}$  & $0.00\,\pm\,0.30$  & $0.00\,\pm\,0.30$                    & $0.00$                             &       \emph{0.00}                               &                    \\
$\epsilon$(C)           & $8.12\,\pm\,0.30$  & $8.20\,\pm\,0.30$                    & $8.20\,\pm\,0.40$                  & $8.23\,\pm\,0.04$\footnotemark[3]    &                    \\
$\epsilon$(N)           & $8.37\,\pm\,0.40$  & $8.26\,\pm\,0.40$                    & $8.18\,\pm\,0.40$                  & $8.15\,\pm\,0.05$\footnotemark[4]    &                    \\
$\epsilon$(O)           & $9.08\,\pm\,0.30$  & $8.93\,\pm\,0.30$                    & $8.93\,\pm\,0.40$                  &       \emph{8.93}                               &                    \\
$^{12}C/ ^{13}C$         & $9\,\pm\,2$        & $10\,\pm\,2$                         & $5\,\pm\,3$                        &        \emph{10}                              &                    \\
$\theta_\mathrm{d}$      & $13.03\,\pm\,0.06$ & $12.34\,\pm\,0.06$                   & $16.43\,\pm\,0.08$                 & $11.00\,\pm\,0.05 $   &                    \\
distance                 & $61.12\,\pm\,2.84$ & $67.48\,\pm\,3.78$                   & $61.08\,\pm\,2.69$                 & $90.80\,\pm\,18.16$\footnotemark[5]  &                    \\
$M_{g}$                  & $2.49\,\pm\,1.48$  & $2.69\,\pm\,1.61$                    & $1.94^{+4.27}_{-1.34}$             & $1.5$                                &                    \\
$\mathrm{A}_\mathrm{v}$  & $0.06\,\pm\,0.15$  & $0.06\,\pm\,0.16$                    & $0.03\,\pm\,0.15$                  & $0.20\,\pm\,0.17$                    &                    \\

\hline
\end{tabular}
\end{minipage}
\flushleft
\footnotesize{
$\qquad \qquad$
$^1$ \citet{McWilliam1990},
$^2$ \citet{Judge1991},
$^3$ \citet{Tsuji1991},
$^4$ \citet{Aoki1997},
$^5$ \citet{Ochsenbein1999}
}
\end{table*}


\section{Discussion} \label{discussion}

\subsection{Proof for a significant flux excess at
  $1.2$\,mm} \label{proofexcess}

Fig.~\ref{sed} shows clear indications for a flux excess at millimeter
and centimeter wavelengths. To prove the flux excess, both the
observational and theoretical uncertainties in the atmosphere models
should first be investigated.

\paragraph{Observational uncertainties}
The uncertainties on the observations in the millimeter/centimeter
wavelength region are typically of the order of $15\,\%$. The IRAS-PSC
error bars given in the catalogue are the statistical 1$\sigma$
uncertainty values; realistic absolute calibration uncertainties are
lacking for the PSC, but are estimated to be ~20\,\% (\emph{D. Kester,
  priv. comm.}). This higher uncertainty was already clear from a
comparison between the Infrared Space Observatory - Short Wavelength
Spectrometer (ISO-SWS) data and the IRAS-PSC and IRAS-LRS data
\citep{VanMalderen2004A&A...414..677V}. Therefore we have used an error
bar of $20\,\%$ on the IRAS-PSC data in our analysis. The IRAS-PSC
fluxes are also colour corrected.\\
The uncertainties on the near- and mid infrared photometry were taken
from the catalogs mentioned in Sect.\,\ref{datareduction}.

\paragraph{Theoretical uncertainties}
As described in \citet{Decin2007A&A...472.1041D}, the uncertainty on
the FIR continuum flux predictions mainly arise from uncertainties on
\emph{(1.)} the estimated stellar temperature and \emph{(2.)} the
neglect of some physical processes.

\emph{(1.)} In the FIR, the dominant continuous opacity arises from
H$^-$ free-free absorption, whose absorption coefficients are nowadays
known at an accuracy of about 1\,\% for wavelengths beyond 0.5\,$\mu$m
over the temperature range between 1\,000 and 10\,000\,K
\citep{Decin2007A&A...472.1041D}. An uncertainty in the estimated
stellar temperature may give rise to an uncertainty on the continuum
predictions of up to 4\,\% for A-M giants.

\emph{(2.)} Since we are tracing regions high up in the atmosphere,
density inhomogeneities and patchy temperature structures may
occur. This kind of 3-dimensional structures are not dealt with in the
1-dimensional {\sc marcs} model atmosphere code. Luckily, the
wavelength regions of interest are in the FIR, where the sensitivity of
the Planck function to the temperature is small. Another important
physical process not included in the {\sc marcs} atmosphere code is
the presence of circumstellar dust and/or a chromosphere or ionized
wind. While the latter is the topic of this study, the first excess
can be excluded from the detailed analysis of the ISO-SWS data for 7
targets in our sample \citet{Decin2003d} and
\citet{VanMalderen2004A&A...414..677V}. \object{$\iota$~Aur} and
\object{$\sigma$~Lib} were not observed by ISO, but good-quality
IRAS-LRS data exist for both objects. The IRAS-LRS data show no sign
of flux excess due to circumstellar dust.

A remark concerning the angular diameters that were used to compute
the fundamental parameters for the {\sc marcs} models is in place here.
The angular diameters are computed from Selby or TCS K-band
photometry. For \object{$\beta$\,Umi}, we have used the Johnson K-band magnitude
of $-1.22$ (Faucherre et al. 1983), yielding a magnitude of $-1.276$ in
the Selby system. Zeropoints are calculated using the Kurucz theoretical
spectrum of Vega, taking into account the observed near-IR excess of
Vega (Absil et al. 2006). For the Selby photometric system we obtain a
zeropoint of $4.0517\,10^{-10}$\,W/m$^2$/$\mu$m, for TCS
$4.4506\,10^{-10}$\,W/m$^2$/$\mu$m. An uncertainty of 0.01\,mag in the K-band
photometric data (0.1\,mag for \object{$\beta$\,Umi}) is propagated in the computation of the uncertainty on
the angular diameter.
At every wavelength, the observed angular diameter represents the
apparent diameter of the stellar surface where
$\tau_{\lambda}\,\sim\,1$. Since at the longer wavelengths, we are
tracing layers that lie further and further outwards, it is expected
that the angular diameter increases with increasing wavelength. If the
angular diameter in the millimeter and centimeter area is considerably
larger than the assumed value, the {\sc marcs} models will
underestimate the flux in this wavelength area.

  To investigate this, we derived the change in height of the
  continuum forming layers with increasing wavelength for the M0
  giant \object{$\beta$~And}. The layer where $\tau_{ross} = 1$ (with
  $\tau_{ross}$ the Rosseland optical depth) defines the stellar radius,
  being in case of \object{$\beta$~And} $R_{\ast}= 6.12\,10^{12}\,$\,cm. The
  flux at $100\,\mu$m is formed at $R_{\ast} + 6.90\,10^{10}\,$\,cm,
  for $150\,\mu$m at $R_{\ast} + 7.89\,10^{10}\,$\,cm and for
  $200\,\mu$m at $R_{\ast} + 1.70\,10^{11}\,$\,cm. From the {\sc
    marcs} model we derive that the flux at $7$\,cm (this is the
  longest wavelength for which we have observations) is formed at
  $R_{\ast} + 2.23\,10^{11}\,$cm, which corresponds to an increase in
  radius of $3.64\,\%$. For the other sample stars, comparable numbers
  are found.  This increase in angular diameter is insufficient to
  explain the observed excess.

  In general, the uncertainties on the theoretical flux
  predictions are in the order of 5 to 10\,\%, excluding the effects
  of a chromosphere or ionized wind. The observed flux excesses are
  hence not caused by inaccuracies in the modeling, but are due to
  physical processes in the stars.


\subsection{Brightness temperature} \label{brightnessT}

\begin{figure*}
\sidecaption
\includegraphics[width=12cm]{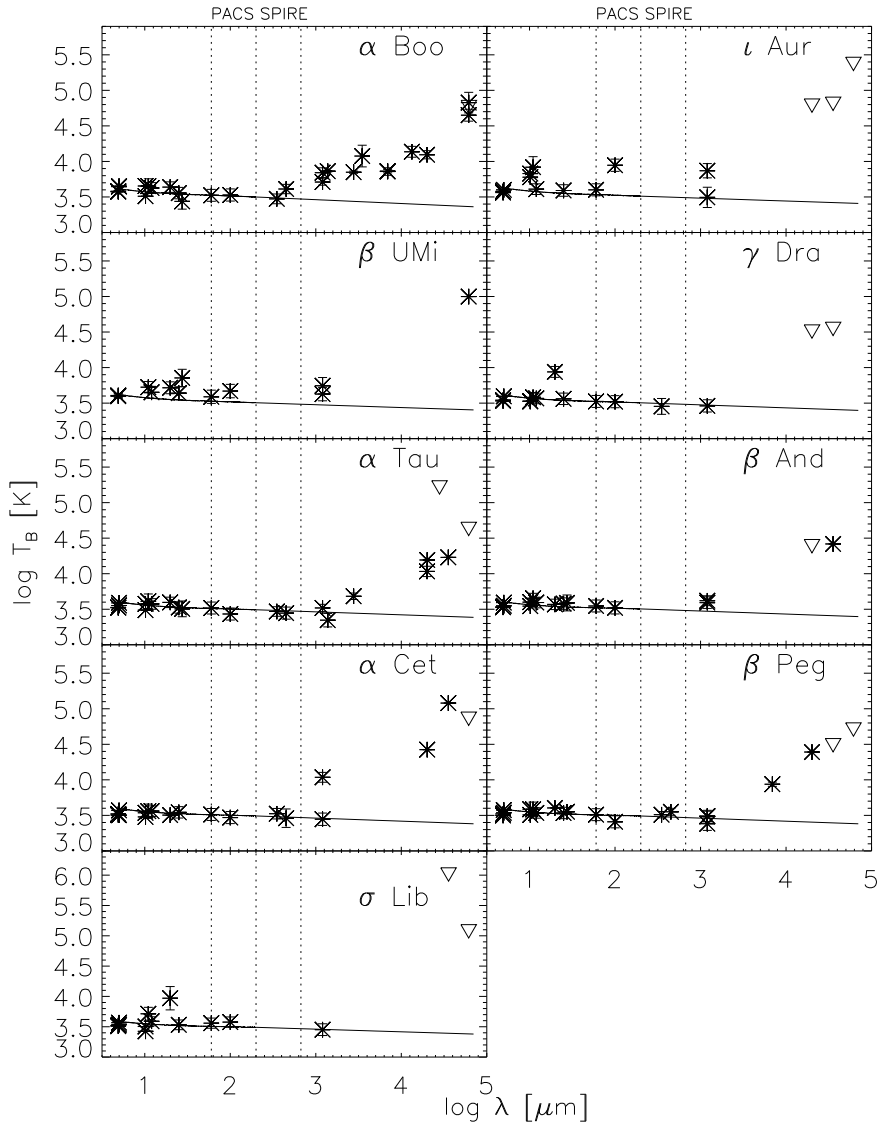}
\caption{Brightness temperature $T_B$ in function of the wavelength
  for the $9$ sample stars between $5\,\mu$m and $7\,$cm. The full
  line indicates the brightness temperatures derived from the {\sc
    marcs} model, the asterisks show the brightness temperatures
  derived from the observations. A reversed triangle represents an
  upper limit. The error bars on the observational data take
  the uncertainty on the observations and on the angular diameter into
  account. The PACS and SPIRE wavelength ranges are indicated by a
  dotted line, to facilitate comparison to the wavelength region were
  a flux excess is seen. As an illustration of the influence of the effective
  temperature, three models with different $T_{\rm{eff}}$ are shown for \object{$\beta$\,UMi}.
  Fig.\,\ref{tbrightbetaumi} gives a clearer view of these models. }
\label{tbright}
\end{figure*}

Fig.~\ref{tbright} provides another window at studying the flux
excess. It shows the brightness temperature over
the full 5\,$\mu$m to 7\,cm wavelength range. The brightness
temperature is defined as the temperature of a black body that gives
the same flux as the model atmosphere at the indicated wavelength, and
can be written as \citep{cohen2005}
\begin{eqnarray} \label{eqtbright}
  T_B(\lambda) = \frac{14387.75/\lambda}{\textrm{ln}\left(
      1+\frac{733.4090\,\theta_D^2}{F_{\nu}(\lambda)\lambda^3}\right)}  \,,
\end{eqnarray}
where $F_{\nu}$ is the observed flux in Jy, $\theta_D$ the angular
diameter in milliarcseconds, $\lambda$ the wavelength in $\mu$m, and
$T_B(\lambda)$ the brightness temperature in K.

\begin{figure}
\sidecaption
\includegraphics[width=9cm,height=6.5cm]{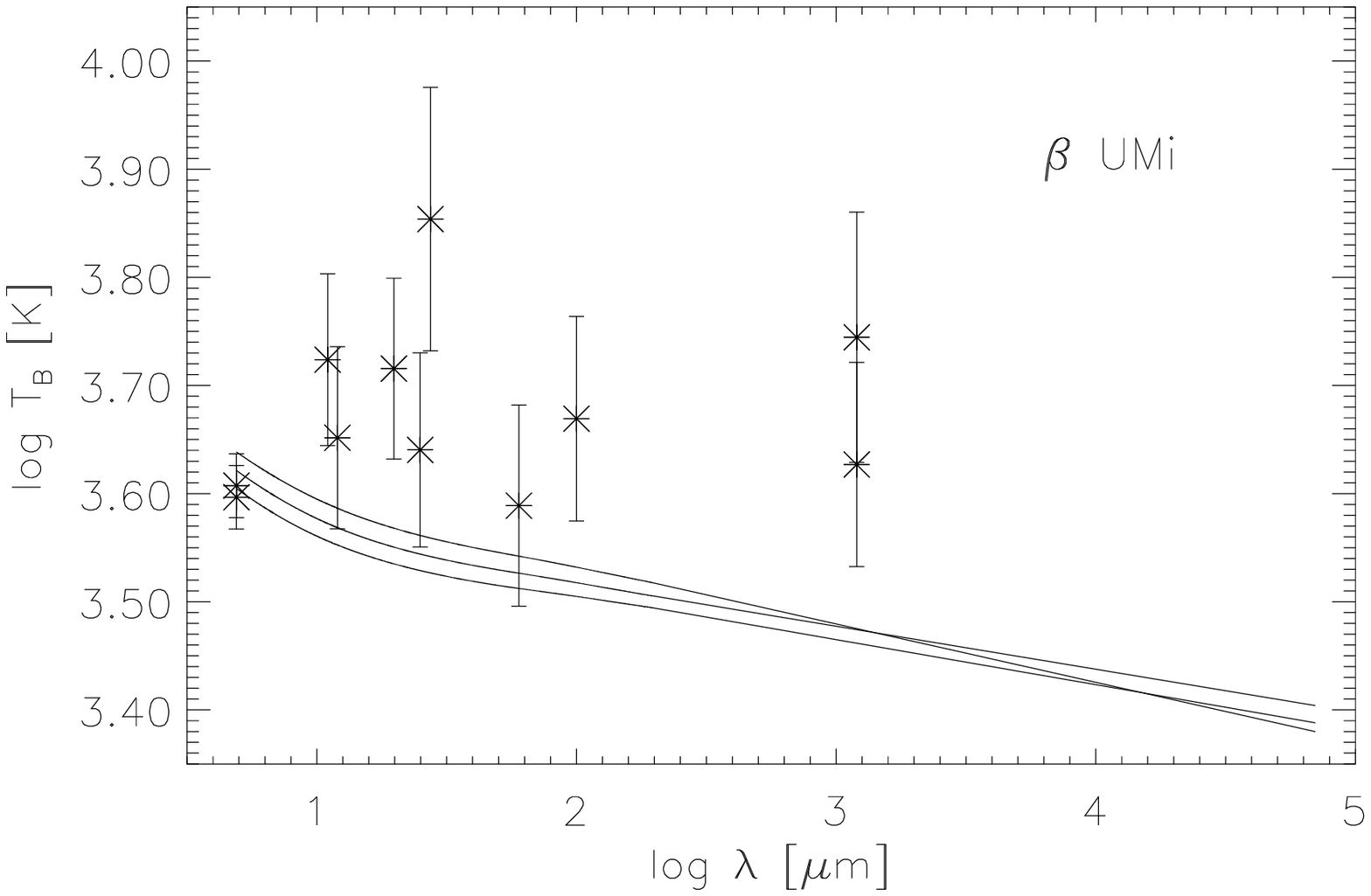}
\caption{As an illustration of the influence of the effective
  temperature, three models with different $T_{\rm{eff}}$ are shown for \object{$\beta$\,UMi}. The effective temperature equals $3885$, $4085$ or $4285$\,K. A different scale then in Fig.\,\ref{tbright} is used to enhance the region where the models differ. The observation at $6.14\,$cm is no longer visible here.}
\label{tbrightbetaumi}
\end{figure}

For comparison also the brightness temperatures from the theoretical
models are plotted in Fig.~\ref{tbright}.  The uncertainty on the data
and on the angular diameter (see previous section) has been propagated
to determine the error bars on the observational brightness
temperatures. The differences between the theoretical brightness temperatures (as derived from the
{\sc marcs} predictions) taking the angular diameter uncertainty into
account are indiscernible. The uncertainty on the effective
temperature has the highest influence on the theoretical brightness
temperature predictions. This is illustrated in case of $\beta$ UMi in
Figs.~\ref{tbright} and \ref{tbrightbetaumi}. In the approximation for
long wavelengths, the formula for the brightness temperature shows
that $T_B(\lambda) \propto 1/\theta_D^2$. Since the angular diameters
at long wavelengths might be underestimated when an extra component
besides the photosphere is present, the brightness temperatures in
Fig.\,\ref{tbright} should be regarded as
upper limits.

In Fig.\,\ref{tbright} one can clearly see the wavelength region where
the flux excess starts for each of the sample stars.
For \object{$\beta$\,UMi}, the observation at $60\,\mu$m still agrees with
the model within the error. Moreover, also the IRAS-LRS and ISO-SWS
data of $\beta$ UMi agree with the model predictions
\citep{VanMalderen2004A&A...414..677V}.  The flux at $100\,\mu$m
is in excess of the model.  The same holds for \object{$\iota\,$Aur}. Note that the IRAS flux at $100\,\mu$m for \object{$\iota\,$Aur} is of lesser quality than the other IRAS points used in our analysis. But recently a weak IR flux excess at 70\,$\mu$m has
been found in the Spitzer-MIPS data \citep{Gordon2007PASP..119.1019G}, which confirms our finding that the excess starts somewhere between $60\,\mu$m en $100\,\mu$m.
For \object{$\alpha$\,Boo}, the model and the data coincide for
wavelengths up until $350\,\mu$m.  The measurement at $450\,\mu$m
lies above the model prediction and the data at longer wavelengths
are all clearly in excess of the predictions.
For \object{$\alpha$\,Cet} the excess seems to start between $450\,\mu$m
and $1.2\,$mm, as the average flux at $1.2\,$mm lies well above the model.
For \object{$\beta$\,And} it starts between $100\,\mu$m and $1.2\,$mm; unluckily
no measurements are available between $100\,\mu$m and $1.2\,$mm.
For \object{$\alpha\,$Tau} and \object{$\beta\,$Peg}, the excess starts at
longer wavelengths, as the fluxes until $1.38\,$mm
(respectively $1.2$\,mm) are still in accordance with the predictions.
For \object{$\gamma\,$Dra} and \object{$\sigma$\,Lib} all available data
coincide with the model predictions, including the measurements at
$1.2$\,mm. However, there are no observational data at longer
wavelengths and the upper limits at centimeter wavelengths are such
that they do not exclude an excess.

It was already argumented that the increase in angular diameter at
longer wavelengths does not lead to significant changes in the
{\sc marcs} predictions. However, possible extra components of the
stellar atmosphere not taken into account by the {\sc
  marcs}-code, such as a chromosphere, can be extended in volume in
comparison to the stellar photosphere. Proof for this kind of
extension can be found in for example \citet{drake1986}. They studied
observations at $2$ and $6$\,cm of, a.o., \object{$\alpha$~Boo}. They
treated the radiation at radio wavelengths as originating from an
optically thick ionized wind (see also Sect.\,\ref{ionizedwind}), for
which they calculated the half-power radius (meaning that half of the
radio emission originates from within this radius). For \object{$\alpha\,$Boo}
this half-power radius at $2$\,cm corresponds to the stellar radius,
but at $6$\,cm this radius had increased to $\simeq 1.7\,R_{\ast}$. In the
next sections, we will elaborate on the possibility that the flux
excess arises from the presence of a chromosphere (corona) or from thermal
emission in an ionized stellar wind.


\subsection{Influence of a chromosphere at NIR wavelengths}\label{wiedemann}

In 1994, Wiedemann et al.\ \nocite{wiedemann1994} studied the
fundamental vibration-rotation lines of CO (at $\sim$4.6\,$\mu$m) in a
set of late-type stars. The CO $\Delta v = 1$ lines are remote sensors
for the thermal conditions in the outermost layers of the
atmosphere. In particular, the strongest CO $\Delta v = 1$ lines occur
at or above the temperature minimum in chromospheric solar and stellar
models.  \citet{wiedemann1994} found observational evidence in favor
of a `thermal bifurcation' model for the atmospheres of their sample
stars.  It consists of two distinct physical phases that co-exist at
chromospheric altitudes. One component is controlled by molecular
cooling and is represented by a radiative equilibrium model atmosphere
with CO induced temperature depression. The second component is
chromospheric and features a temperature inversion produced by the
deposition of mechanical energy. Any observed spectrum from an
atmosphere with thermal bifurcation is to be interpreted as a spatial
sum over the two types of thermal regions with appropriate geometrical
weighting factors.

\citet{wiedemann1994} concluded that the observed infrared CO $\Delta
v =1 $ spectrum at 4.6\,$\mu$m of one group of stars, containing
\object{$\alpha$\,Boo}, \object{$\alpha$\,Tau} and
\object{$\gamma$\,Dra}, is well described by homogeneous radiative
equilibrium models. The near-IR CO spectra for this group of so-called
`quiet' stars indicate that the cool regions dominate the stellar
surface for heights between $\sim$ $10^{-1/2}$ and $10^{-2}$ g/cm$^2$
in mass column density, and have large filling factors. These stars
are said to have a `COmosphere'. For a second group of stars, the CO
$\Delta v = 1$ spectrum is poorly represented by radiative equilibrium models,
and is compatible with a chromosphere covering the stellar surface
homogeneously at these altitudes. After investigation of different
chromospheric indicators, it also became clear that the stars in the
first group show only little chromospheric activity.

Three stars from our sample, \object{$\alpha$\,Boo}, \object{$\alpha$\,Tau} and
\object{$\gamma$\,Dra}, belong to the first group of 'quiet stars'\ i.e.\ their
CO $\Delta v = 1$ spectrum indicates that the COmosphere dominates the
thermal structure at heights between $\sim$ $10^{-1/2}$ and $10^{-2}$
g/cm$^2$ in mass column density. For these stars, the spectrum at NIR wavelengths
is not influenced by their chromospheric activity. In the following sections, we
investigate if this remains true at longer wavelengths.


\subsection{Coronal, transition region and wind dividing lines} \label{dividing lines}

\citet{linsky1979} introduced a dividing line (further denoted by DL)
in the cool part of the HR diagram on the basis of ultra-violet spectra of
late-type stars. Stars to the blue side of the `transition region DL'
were termed `solar-type', as they showed spectral lines formed at
temperatures of $5 \times 10^3 - 2 \times 10^5$\,K, indicative of
chromospheres, transition regions and by implication unseen coronae at
even hotter temperatures. Stars to the red side are of `non-solar type':
they only exhibited lines formed at temperatures below $10\,000 -
20\,000$\,K, indicative of chromospheres only. \citet{ayres1981}
attempted to observe soft X-ray emission from late-type stars, being a
signature of stellar coronae (T $>$ $10^6$\,K). They found an `X-ray
DL' roughly coinciding with the `transition region DL' from
\citet{linsky1979}; only stars to the blue side were detected in
X-rays. \citet{stencel1980} studied the morphology of \ion{Mg}{ii}\,h
and k resonance lines and they also found a similar distinction in
position in the HR-diagram between stars with a low-velocity wind in
their chromosphere (to the red side of the line) and stars without (to
the blue side).

\begin{figure}
\includegraphics[width=9cm,height=6.5cm]{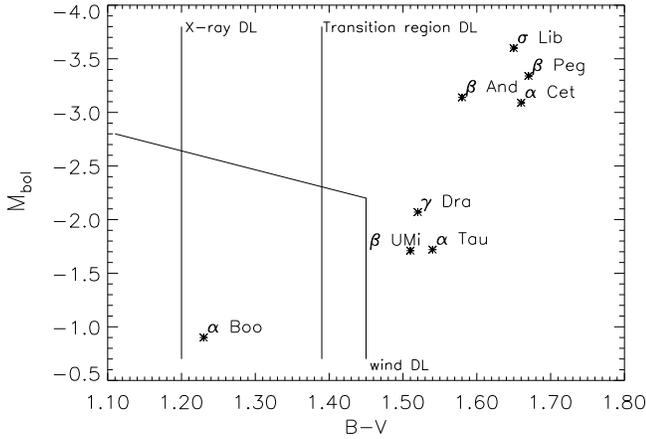}
\caption{The position of the different DLs in the HR diagram, together
  with the eight luminosity class III giants from our sample. The `X-ray
  DL' is derived by \citet{hunsch1996}, the `transition region DL' by
  \citet{haisch1990}, and the `wind DL' by \citet{reimers1996} (see
  text for more details).}
\label{dividingline}
\end{figure}

Since the first discovery of the DLs, several authors have confirmed
the dichotomy in late-type giant atmospheres, but more sensitive
observations have also revised the location of the different DLs in
the HR diagram. \citet{hunsch1996} place the `X-ray DL' at
$B-V\,=\,1.2$ for luminosity class III giants. According to
\citet{reimers1996}, the `wind DL' runs vertically at
$B-V\thickapprox1.45$, for $B-V < 1.45$ it runs nearly horizontal to
$B-V\thickapprox1.0$ at $M_{bol}\thickapprox-2.8$.  And
\citet{haisch1990} found that the latest occurrence of emission lines
from \ion{C}{iv} or \ion{Si}{iv}, indicative of a transition region,
occurs at K4 III, corresponding to $B-V\,=\,1.385$
\citep{Gray1992oasp.book.....G}. This places all of our eight stars
with luminosity class III in the category of late-type stars with a
chromosphere and a cool wind (see Table~\ref{spectraltype} and
Fig.\,\ref{dividingline}).  \object{$\alpha\,$Boo}, although
located to the blue side of the wind DL, is known to posses a cool wind
\citep{ayres1982} and is regarded as an archetype of a non-coronal
star. The K2~II giant \object{$\iota\,$Aur} was detected
in X-rays by \citet{reimers1996} and shows \ion{C}{iv} and
\ion{Si}{iv} emission \citep{haisch1990}, but also circumstellar
\ion{Ca}{ii}\,H+K lines \citep{reimers1996}. It is a so-called
`hybrid' giant. Hybrid giants are located to the red (i.e.\
non-coronal side) of the `transition region DL', but show the
existence of both transition region plasma and large mass-loss rates
($10^{-9}$ -- $10^{-10}$\,M$_{\odot}$/yr)
\citep{harper1992}.


It should, however, be noted that the distinction between coronal and
non-coronal giants is not so clear-cut. In a number of recent
articles, evidence is given that all giants show some level of X-ray
emission when observed with sufficient sensitivity. On the other hand,
all observations confirm the significant drop in X-ray emission around
early K spectral type for luminosity class III.

X-ray emission from cool stars is linked to the confinement of hot coronal
matter in magnetic structures. An understanding of the origin of these
magnetic structures is closely related to the physical explanation for
the existence of the DLs, both of which are still under
discussion. Most authors seem to agree on the fact that a change in
surface magnetic field topology is responsible for the existence of
the DLs. According to \citet{rosner1995}, the field topology of a red
giant changes from a large scale organized and closed configuration
binding coronal gas, to a largely open magnetic field giving rise
to a massive cool wind, as the star
evolves along the RGB from the left of the DLs to the right. The
transition in topology is ascribed to a change in the origin of the
field: as the stellar rotation drops below a critical value, the
spin-catalyzed dynamo gives way to a field generation mechanism
requiring fluid turbulence as found in the convection
zone. It has also been
suggested that the magnetic flux tubes that rise up from underneath
the convection zone to the stellar surface where they form large scale
coronal loops, become trapped in the convective envelope as the
convection zone deepens to the right of the DLs
\citep{holzwarth2001}. However, \citet{ayres1997, ayres2003} have
found evidence in ``non-coronal'' giants that coronal loops can still
rise to the stellar surface: the loops extend beyond the cold
molecular layer just above the stellar photosphere, but are at least
partially submerged in the chromosphere/COmosphere, where the coronal
X-rays are attenuated by overlying material. It is also still unclear
if stars evolving along the RGB cross the DLs \citep[as was
a.o. postulated by][]{rosner1995}), or if the evolution tracks run
parallel to them \citep{hunsch1996}. In this last scenario, the
difference in X-ray activity on either side of the DLs would be due to
a different rotational history of each star implying a difference in
spin-catalyzed dynamo strength.

In the following sections we will investigate if the observed flux
excess in the (sub)millimeter and centimeter wavelength range in our selected
late-type giants can be explained by the presence of a chromosphere
(corona) and/or by thermal emission from an ionized stellar wind. As we
shall show in Sect.\,\ref{ionizedwind} and Sect.\,\ref{chromosphere},
the radiation coming from a chromosphere or from an ionized wind
exhibits a different frequency dependence $F_{\nu} \varpropto
\nu^{\alpha}$. The spectral indices $\alpha$ as determined from the
available observations will be compared to the theoretical predictions
to determine the cause of the flux excess.


\subsection{Thermal emission from an ionized wind} \label{ionizedwind}
Stars with an ionized wind emit an excess of continuum emission at
long wavelengths, i.e. from the IR to the radio region, in addition to the black
body flux emission. This excess
flux is due to free-free emission or Bremsstrahlung from the wind. To
derive the wavelength dependence of the thermal emission from an
ionized wind, we will use a model from \citet{olnon1975}. This article
gives analytic expressions for the flux originating from a stellar
wind, assuming a homogeneous, spherical geometry with a uniform
electron temperature and with \ion{H}{ii} as the only constituent. In
reality, the hydrogen in the winds of these late-type stars will be
only partially ionized \citep{drake1987}. The free-free absorption
coefficient per unit mass $\kappa_{\nu}^{ff}$ in cm$^2$ g$^{-1}$ of an
ionized gas at long wavelengths is \citep[e.g.][]{lamers1999}
\begin{eqnarray} \label{eqkappanuff}
  \kappa_{\nu}^{ff}=1.78\,10^{-2}\,Z^2\,\nu^{-2}\,g_{\nu}\,T^{-3/2}\,\frac{n_i\,n_e}{\rho}
\end{eqnarray}
where $Z^2$ is the square of the charge of the atoms, $n_e$ and $n_i$
are the electron and ion densities in $\textrm{cm}^{-3}$, $\rho$ is
the density in $\textrm{g\,cm}^{-3}$  and $g_{\nu}$ is the gaunt
factor which will be approximated by a power law
\begin{eqnarray}\label{eqgauntfactor}
   g_{\nu} \simeq 1.37\, T^{0.135}\,\lambda^{0.084}
\end{eqnarray}
where $\lambda$ is expressed in cm.
It is clear from these expressions that the wavelength dependence of
the emitted flux will be the same, regardless of the main contributor
to the flux, be it \element[-][][][]{H}ff or \ion{H}{i}\,ff.

Model~V from \citet{olnon1975}, the truncated power law distribution
is of particular intrest here. It assumes an electron density distribution
with a homogeneous sphere in its centre: $n_e
\varpropto r^{-2}$ for $r \geq R$, for $r \leq R$ $n_e$
is constant. Using Eq.~(\ref{eqkappanuff}) in his expressions, the
model predicts $F_{\nu} \varpropto \nu^{0.611}$ in the optically thick
limit and $F_{\nu} \varpropto \nu^{-0.084}$ in the optically thin
limit. These approximations can be generalised to density
distributions $n_e \varpropto r^{-\beta}$ with $\beta$ $>$ $1.5$. A
value for $\beta$ differing from $2$ can be caused by a non-constant
velocity distribution in the wind. This is a very plausible
explanation if the radio emission originates from the wind
acceleration zone. A decreasing (increasing) fractional ionisation
rate (this is the number of free electrons per neutral hydrogen atom)
with radius can also lead to a higher (lower) value for $\beta$. An
extreme case of this scenario would be the existence of an outer
cutoff radius $r_0$ to the ionized portion of the wind. Although
doubtful for the targets in our study, this cutoff might be
caused for example by the formation of dust at this location
in the wind, `quenching' the ionized material \citep{drake1986}. The spectral index would
change for $\nu$ $<$ $\nu_0$, where $\nu_0$ is determined by the
cutoff radius. For K to mid M cool wind giants, most estimates of the
dust-formation region (if present at all) indicate $r_0/R_{\ast}
\approx 10$. It can be shown that the spectrum is only influenced by
this transition for $\lambda \gtrsim 30\,$cm \citep{drake1986}. No
observations beyond this wavelength are being used in this article,
hence such a spatial restriction of the ionized region is of no
importance for our discussion.

In the optically thin case the wavelength dependence is not influenced by the value for $\beta$. In the optically thick case we have
\begin{eqnarray}
 F_{\nu} \propto \nu^{\alpha}\ \ \ \textrm{with} \ \ \ \alpha = \frac{2\,\beta - 3.084}{\beta - 0.5} \,.
\end{eqnarray}
For $\beta=1.5$ we find $\alpha=-0.084$, which is the same frequency
dependence as in the optically thin case. $\alpha$ goes asymptotically
to $2$ as $\beta \rightarrow +\infty$, but we do expect $\beta$ to
fluctuate around $2$.


\subsection{Chromospheric emission}\label{chromosphere}
The continuum radiation from a chromosphere will be mainly free-free
emission from \element[-][][][]{H} and \ion{H}{i}. The flux can be
written as
\begin{eqnarray}
 F_{\lambda} = \kappa_{\lambda}^{ff} \rho B_{\lambda}(T) V
\end{eqnarray}
in the optically thin case, where V is the volume of the emitting
region and $B_{\lambda}$ is the Planck function. This expression can
be derived from Eq.~(1) in \citet{olnon1975}, see
also \citet{skinner1987}. With the use of Eq.~(\ref{eqkappanuff}) and
the Rayleigh-Jeans approximation for the Planck function where
$B_{\lambda}(T) \propto \lambda^{-4}$, we find a wavelength dependence
of $F_{\lambda} \propto \lambda^{-1.916}$. In case of an optically
thick chromosphere, we are looking at a black body with a temperature
equal to the electron temperature at the layer where $\tau_{\lambda} =
1$. In that case we find $F_{\lambda} \propto \lambda^{-4}$. In the
above approximations, the chromosphere is treated as a homogeneous
region, with uniform densities and electron temperature.

The model above is also applicable to thermal emission from a
corona. This implies that the same wavelength dependence will be found
for corona and chromosphere and that our analysis will not allow to
discriminate between these two sources of free-free emission.


\subsection{Spectral indices for our program stars} \label{spectralindex}

\begin{figure*}
\sidecaption
\includegraphics[width=12cm]{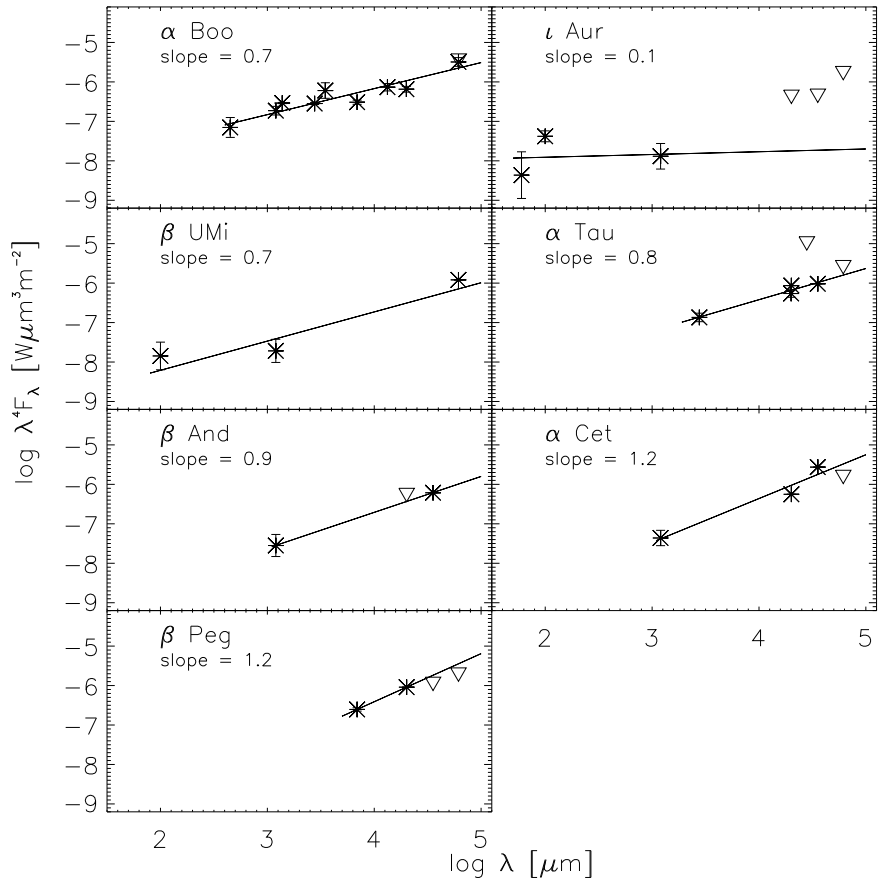}
\caption{These figures show the true flux excess for the $7$ stars in
  our sample for which a clear flux excess is detected (i.e.\
  \object{$\gamma$~Dra} and \object{$\sigma$ Lib} are excluded, since
  only upper limits are available). The observations (with the
  continuum already subtracted) are represented by asterisks. The
  full lines show the linear least-squares fit to the data. If there
  were different measurements at one wavelength, the fluxes were
  replaced by a weighted average. Error bars are shown, but they often fall
  within the symbols for the fluxes. A triangle represents an upper
  limit. The slope for each target is indicated in the upper corner of
each panel. }
\label{helling}
\end{figure*}

Fig.\,\ref{helling} shows the true flux excess (the
theoretically calculated flux is already subtracted from the data) at
millimeter/centimeter wavelengths. We have chosen to plot $\lambda^4
F_{\lambda}$ in function of wavelength on a logarithmic scale, since
the data then follow a horizontal line in case of an optically thick
chromosphere. To determine the spectral indices, we searched for the
best fitting line through our data using a least-squares method. The starting
point for these lines coincides with the first wavelength where a flux
excess is noticeable (see Sect.\,\ref{brightnessT}), except for
\object{$\iota\,$Aur} (see later in this section). For the
least-squares fit we did not take any upper limits into
account. Table\,\ref{indices} summarizes which spectral indices are
expected for a chromosphere and for an ionized wind based on the
simplified analytic expressions derived above.

The least-squares fits for \object{$\alpha$\,Cet} and \object{$\beta$\,Peg} have slopes of $1.2$.
This value lies closest to the spectral index expected for an optically thick ionized wind, with
$\beta$ equal to $2.2$. For \object{$\beta\,$And}, \object{$\alpha\,$Tau}, \object{$\beta$\,UMi}
and \object{$\alpha$\,Boo}, the slopes have values of respectively $0.9$, $0.8$, $0.7$ and $0.7$.
These values lie somewhere in between the slope of an optically thick ionized wind (with $\beta$
equal to respectively $2.8$, $3.1$, $3.5$ and $3.5$) and an optically thick chromosphere.
For \object{$\iota$\,Aur} only two fluxes are available in excess of the model, among which the
IRAS flux at $100\,\mu$m which is of lesser quality. A least-squares fit to these two points leads
to a slope of $-0.5$. For a better determination of the slope, we can include the IRAS flux at
$60\,\mu$m, which is of normal quality and just barely agrees with the model within the error.
This fit gives a slope of $0.1$ (see Fig.\,\ref{helling}). Hence, only for \object{$\iota$\,Aur},
an optically thick chromosphere is found.

For all stars except \object{$\iota$\,Aur} and \object{$\alpha$\,Boo},
the least-squares fits are made to data at wavelengths longer than
$1.2$\,mm, because only from this wavelength onwards a flux excess was
detectable. In these cases, an optically thick ionized wind is the
most likely explanation for the observed excess, although the density
distributions sometimes show quite large deviations from the average
$n_e \varpropto r^{-2}$. The only star where an optically thick
chromosphere is seen, the hybrid giant \object{$\iota$\,Aur}, has a
line fitted only to wavelengths shorter than $1.2\,$mm, because the
flux excess was already present in this region and no measurements at
longer wavelengths were available. We therefore propose that at shorter wavelengths
($\lambda \lesssim 1\,$mm) an optically thick chromosphere is being
sampled and at longer wavelengths the continuum forming layers lie
further outwards in the atmosphere and the observations show an
optically thick ionized wind. It is very well
possible that for wavelengths slightly longer than $\sim$1\,mm, the
optically thick chromosphere is still visible, as this would explain
the deviating values for $\beta$ in the optically thick ionized winds.


\renewcommand{\arraystretch}{1.3}
\begin{table}
\begin{center}
\begin{tabular}{|l|c|c|}
\hline
        &  optically thin   &    optically thick    \\  \hline
chromosphere & $\lambda^4F_{\lambda} \propto \lambda^{2.1}$ & $\lambda^4F_{\lambda} \propto \lambda^0$\\
ionized wind & $\lambda^4F_{\lambda} \propto \lambda^{2.1}$ & $\lambda^4F_{\lambda} \propto \lambda^{1.4}$\\
\hline
\end{tabular}
\end{center}
\caption{Summary of the wavelength dependencies of the flux
  derived for simplified models for the chromosphere and the ionized
  wind. The table gives the wavelength
  dependence for an optically thick ionized wind with a density
  distribution $n_{e} \propto r^{-\beta}$ with $\beta = 2$. For $\beta
  = 1.5$ we have $\lambda^4F_{\lambda} \propto \lambda^{2.1}$ and for
  $\beta \rightarrow \infty$ we have $\lambda^4F_{\lambda} \propto
  \lambda^0$.}
\label{indices}
\end{table}


\section{Conclusions}\label{conclusions}
In $7$ out of the $9$ K- and M-giants examined, a clear flux excess at
millimeter and/or centimeter wavelengths was found, for the other two
targets only observational upper limits are available at centimeter
wavelengths. The selected
stars have low chromospheric and coronal activity and three of them
do belong to the group of so-called `quiet' K- and M-giants,
where the near-infrared CO $\Delta v = 1 $ lines indicate that the
CO-cooled regions, as predicted by radiative equilibrium models,
dominate over the chromosphere at altitudes between $\sim$ $10^{-1/2}$
and $10^{-2}$ g/cm$^2$ in mass column density. On the basis of this
study, it seems that for these stars the homogeneous atmosphere models
based on radiative equilibrium are able to reproduce the CO spectrum
around $4.6\,\mu$m, but clearly fail to reproduce the flux at
millimeter and centimeter wavelengths. At these far-IR wavelengths,
the presence of a chromosphere and ionized stellar wind cause a clear
flux excess.

The observed excess at wavelengths shorter than $\sim$1\,mm is most
likely to be attributed to an optically thick chromosphere, where an
optically thick ionized wind is being sampled at longer
wavelengths. The wavelength region where the excess starts depends
upon the star under consideration. The most extreme cases are
\object{$\iota$\,Aur} and \object{$\beta$\,UMi}, where the excess
starts somewhere between $60\,\mu$m and $100\,\mu$m.  These findings
have implications for the roles of
these standard stars as fiducial calibrators for PACS (wavelengths
between $60$ and $210\,\mu$m) and SPIRE (between $200$ and
$670\,\mu$m). For \object{$\alpha$~Boo} the flux
excess is already present at the SPIRE (but not at the PACS)
wavelengths. For \object{$\alpha$~Cet} it might be present at SPIRE
wavelengths and for \object{$\beta$~And} the excess might already
start at the PACS wavelengths, but a lack of observations in these
regions makes it impossible to indicate the precise start of the
excess. \object{$\alpha$~Tau}, \object{$\beta$~Peg},
\object{$\gamma$~Dra} and \object{$\sigma$~Lib} show no flux excess in
the PACS and SPIRE range, but especially for \object{$\sigma$\,Lib}, only few observations are
available in the relevant region. \object{$\iota$~Aur} and \object{$\beta$\,UMi} show a clear
flux excess from $100\,\mu$m onward, and should not be used as a
calibrator beyond $60\,\mu$m.



\begin{acknowledgements}
This work is based on observations collected at the European Southern
Observatory, La Silla, Chile within program ESO 71.D-0600 and on observations
collected with the IRAM 30m telescope within project $038\_03$. We would like
to thank R. Zylka and S. Leon for their support for the data reduction.
The research at the Caltech Submillimeter Observatory is supported by grant
AST-0540882 from the National Science Foundation.

SD and LD acknowledge financial support from the Fund for Scientific
Research - Flanders (Belgium).\\

\end{acknowledgements}


\bibliographystyle{aa}
\bibliography{dehaes}

\Online

\begin{appendix}
\section{Summary of the photometric data used in this study}

\renewcommand{\arraystretch}{1.05}
\begin{table*}[b]
\caption{Photometric data used in this study for the targets
  \object{$\alpha$~Boo} and \object{$\iota$~Aur}. The literature
  references are specified at the end of Table~\ref{tablephotometry5}. }
\label{tablephotometry1}
\centering
\begin{tabular}{|l|l|l||l|l|l|}
\hline
\multicolumn{3}{|c||}{$\alpha$\,Boo} & \multicolumn{3}{c|}{$\iota$\,Aur} \\
\hline\hline
 \multicolumn{1}{|c}{$\lambda$} & \multicolumn{1}{|c|}{$\lambda F_{\lambda}$} &  ref. & \multicolumn{1}{|c|}{$\lambda$} & \multicolumn{1}{c|}{$\lambda F_{\lambda}$} & ref. \\
 \multicolumn{1}{|c}{$[\mu\textrm{m}]$} & \multicolumn{1}{|c|}{$[\textrm{W} \textrm{m}^{-2}]$} & & \multicolumn{1}{|c|}{$[\mu\textrm{m}]$} & \multicolumn{1}{c|}{$[\textrm{W} \textrm{m}^{-2}]$} &  \\
\hline
    $3.46\,10^{-1}$  &  $1.22\,10^{-9}\,\pm\,1.77\,10^{-11}$   &   $4$         &    $3.46\,10^{-1}$  &  $4.49\,10^{-11}\,\pm\,2.48\,10^{-13}$   &   $4$         \\
    $3.60\,10^{-1}$  &  $1.62\,10^{-9}\,\pm\,0.00$             &   $1$         &    $3.60\,10^{-1}$  &  $6.23\,10^{-11}\,\pm\,0.00$             &   $1$         \\
    $4.01\,10^{-1}$  &  $4.37\,10^{-9}\,\pm\,6.35\,10^{-11}$   &   $4$         &    $4.01\,10^{-1}$  &  $2.05\,10^{-10}\,\pm\,1.13\,10^{-12}$   &   $4$         \\
    $4.23\,10^{-1}$  &  $7.40\,10^{-9}\,\pm\,9.27\,10^{-11}$   &   $4$         &    $4.23\,10^{-1}$  &  $4.12\,10^{-10}\,\pm\,1.70\,10^{-12}$   &   $4$         \\
    $4.40\,10^{-1}$  &  $1.02\,10^{-8}\,\pm\,0.00$             &   $1$         &    $4.40\,10^{-1}$  &  $6.20\,10^{-10}\,\pm\,0.00$             &   $1$         \\
    $4.48\,10^{-1}$  &  $1.08\,10^{-8}\,\pm\,1.57\,10^{-10}$   &   $4$         &    $4.48\,10^{-1}$  &  $6.48\,10^{-10}\,\pm\,3.58\,10^{-12}$   &   $4$         \\
    $5.39\,10^{-1}$  &  $2.02\,10^{-8}\,\pm\,2.93\,10^{-10}$   &   $4$         &    $5.39\,10^{-1}$  &  $1.55\,10^{-9}\,\pm\,8.55\,10^{-12}$    &   $4$         \\
    $5.49\,10^{-1}$  &  $2.15\,10^{-8}\,\pm\,2.18\,10^{-10}$   &   $4$         &    $5.49\,10^{-1}$  &  $1.69\,10^{-9}\,\pm\,3.12\,10^{-12}$    &   $4$         \\
    $5.50\,10^{-1}$  &  $2.17\,10^{-8}\,\pm\,0.00$             &   $1$         &    $5.50\,10^{-1}$  &  $1.74\,10^{-9}\,\pm\,0.00\,10^{+00}$    &   $1$         \\
    $5.81\,10^{-1}$  &  $2.48\,10^{-8}\,\pm\,3.60\,10^{-10}$   &   $4$         &    $5.81\,10^{-1}$  &  $2.02\,10^{-9}\,\pm\,1.12\,10^{-11}$    &   $4$         \\
    $7.00\,10^{-1}$  &  $3.30\,10^{-8}\,\pm\,0.00$             &   $1$         &    $7.00\,10^{-1}$  &  $2.88\,10^{-9}\,\pm\,0.00$              &   $1$         \\
    $7.00\,10^{-1}$  &  $3.33\,10^{-8}\,\pm\,0.00$             &   $6$         &    $9.00\,10^{-1}$  &  $3.84\,10^{-9}\,\pm\,0.00$              &   $1$         \\
    $9.00\,10^{-1}$  &  $3.77\,10^{-8}\,\pm\,0.00$             &   $1$         &    $1.24$           &  $3.98\,10^{-9}\,\pm\,6.89\,10^{-10}$    &   $12$        \\
    $9.00\,10^{-1}$  &  $3.81\,10^{-8}\,\pm\,0.00$             &   $6$         &    $1.25$           &  $3.31\,10^{-9}\,\pm\,0.00$              &   $1$         \\
    $1.24$           &  $3.08\,10^{-8}\,\pm\,4.45\,10^{-9}$    &   $12$        &    $1.25$           &  $3.40\,10^{-9}\,\pm\,0.00$              &   $6$         \\
    $1.25$           &  $2.88\,10^{-8}\,\pm\,0.00$             &   $1$         &    $1.66$           &  $3.51\,10^{-9}\,\pm\,4.91\,10^{-10}$    &   $12$        \\
    $1.25$           &  $3.28\,10^{-8}\,\pm\,0.00$             &   $6$         &    $2.16$           &  $2.01\,10^{-9}\,\pm\,2.93\,10^{-10}$    &   $12$        \\
    $2.16$           &  $1.35\,10^{-8}\,\pm\,2.11\,10^{-9}$    &   $12$        &    $2.20$           &  $1.53\,10^{-9}\,\pm\,0.00$              &   $1$         \\
    $2.20$           &  $1.30\,10^{-8}\,\pm\,0.00$             &   $1$         &    $2.20$           &  $1.60\,10^{-9}\,\pm\,3.43\,10^{-11}$    &   $7$         \\
    $2.20$           &  $1.39\,10^{-8}\,\pm\,1.53\,10^{-10}$   &   $7$         &    $3.40$           &  $5.65\,10^{-10}\,\pm\,0.00$             &   $1$         \\
    $3.40$           &  $4.92\,10^{-9}\,\pm\,0.00$             &   $1$         &    $3.50$           &  $5.30\,10^{-10}\,\pm\,1.82\,10^{-11}$   &   $7$         \\
    $3.40$           &  $4.83\,10^{-9}\,\pm\,0.00$             &   $1$         &    $3.50$           &  $4.83\,10^{-10}\,\pm\,1.82\,10^{-11}$   &   $7$         \\
    $3.50$           &  $4.33\,10^{-9}\,\pm\,1.03\,10^{-10}$   &   $7$         &    $4.20$           &  $3.80\,10^{-10}\,\pm\,1.40\,10^{-10}$   &   $2$         \\
    $3.50$           &  $4.02\,10^{-9}\,\pm\,1.03\,10^{-10}$   &   $7$         &    $4.90$           &  $1.62\,10^{-10}\,\pm\,5.73\,10^{-12}$   &   $7$         \\
    $4.20$           &  $2.63\,10^{-9}\,\pm\,7.26\,10^{-10}$   &   $2$         &    $4.90$           &  $1.53\,10^{-10}\,\pm\,5.73\,10^{-12}$   &   $7$         \\
    $4.90$           &  $1.41\,10^{-9}\,\pm\,1.56\,10^{-11}$   &   $7$         &    $5.00$           &  $1.65\,10^{-10}\,\pm\,0.00$             &   $1$         \\
    $4.90$           &  $1.38\,10^{-9}\,\pm\,1.56\,10^{-11}$   &   $7$         &    $1.02\,10^{1}$   &  $4.34\,10^{-11}\,\pm\,0.00$             &   $1$         \\
    $5.00$           &  $1.65\,10^{-9}\,\pm\,0.00$             &   $1$         &    $1.02\,10^{1}$   &  $3.81\,10^{-11}\,\pm\,0.00$             &   $6$         \\
    $5.00$           &  $1.68\,10^{-9}\,\pm\,0.00$             &   $6$         &    $1.10\,10^{1}$   &  $4.37\,10^{-11}\,\pm\,1.61\,10^{-11}$   &   $2$         \\
    $1.02\,10^{1}$   &  $1.60\,10^{-10}\,\pm\,0.00$            &   $1$         &    $1.20\,10^{1}$   &  $1.54\,10^{-11}\,\pm\,3.07\,10^{-12}$   &   $3$         \\
    $1.02\,10^{1}$   &  $2.38\,10^{-10}\,\pm\,0.00$            &   $6$         &    $2.50\,10^{1}$   &  $1.76\,10^{-12}\,\pm\,3.53\,10^{-13}$   &   $3$         \\
    $1.10\,10^{1}$   &  $1.91\,10^{-10}\,\pm\,5.27\,10^{-11}$  &   $2$         &    $6.00\,10^{1}$   &  $1.36\,10^{-13}\,\pm\,2.73\,10^{-14}$   &   $3$         \\
    $1.20\,10^{1}$   &  $1.39\,10^{-10}\,\pm\,2.79\,10^{-11}$  &   $3$         &    $1.00\,10^{2}$   &  $6.66\,10^{-14}\,\pm\,1.33\,10^{-14}$   &   $3$         \\
    $1.98\,10^{1}$   &  $3.39\,10^{-11}\,\pm\,6.25\,10^{-12}$  &   $2$         &    $1.20\,10^{3}$   &  $3.25\,10^{-17}\,\pm\,7.50\,10^{-18}$   &   $5$         \\
    $2.50\,10^{1}$   &  $1.39\,10^{-11}\,\pm\,2.78\,10^{-12}$  &   $3$         &    $1.20\,10^{3}$   &  $1.37\,10^{-17}\,\pm\,4.50\,10^{-18}$   &   $10$        \\
    $2.74\,10^{1}$   &  $7.95\,10^{-12}\,\pm\,2.20\,10^{-12}$  &   $2$         &    $2.01\,10^{4}$   &  $<\,6.13\,10^{-20}\,\,\,\,$            &   $5$         \\
    $6.00\,10^{1}$   &  $9.79\,10^{-13}\,\pm\,1.96\,10^{-13}$  &   $3$         &    $3.55\,10^{4}$   &  $<\,1.18\,10^{-20}\,\,\,\,$            &   $5$         \\
    $1.00\,10^{2}$   &  $2.17\,10^{-13}\,\pm\,4.35\,10^{-14}$  &   $3$         &    $6.14\,10^{4}$   &  $<\,8.30\,10^{-18}\,\,\,\,$            &   $5$         \\
    $3.50\,10^{2}$   &  $4.53\,10^{-15}\,\pm\,7.19\,10^{-16}$  &   $13$        &                     &                                          &               \\
    $4.50\,10^{2}$   &  $2.94\,10^{-15}\,\pm\,4.46\,10^{-16}$  &   $13$        &                     &                                          &               \\
    $1.20\,10^{3}$   &  $2.64\,10^{-16}\,\pm\,4.17\,10^{-17}$  &   $9$         &                     &                                          &               \\
    $1.20\,10^{3}$   &  $1.95\,10^{-16}\,\pm\,2.00\,10^{-17}$  &   $5$         &                     &                                          &               \\
    $1.38\,10^{3}$   &  $1.82\,10^{-16}\,\pm\,3.72\,10^{-18}$  &   $8$         &                     &                                          &               \\
    $2.77\,10^{3}$   &  $2.18\,10^{-17}\,\pm\,7.48\,10^{-19}$  &   $8$         &                     &                                          &               \\
    $3.49\,10^{3}$   &  $1.84\,10^{-17}\,\pm\,6.45\,10^{-18}$  &   $5$         &                     &                                          &               \\
    $6.92\,10^{3}$   &  $1.45\,10^{-18}\,\pm\,8.23\,10^{-20}$  &   $11$        &                     &                                          &               \\
    $6.92\,10^{3}$   &  $1.43\,10^{-18}\,\pm\,1.73\,10^{-19}$  &   $11$        &                     &                                          &               \\
    $1.33\,10^{4}$   &  $<\,1.12\,10^{-17}\,\,\,\,$            &   $5$         &                     &                                          &               \\
    $1.33\,10^{4}$   &  $3.82\,10^{-19}\,\pm\,6.74\,10^{-20}$  &   $11$        &                     &                                          &               \\
    $2.00\,10^{4}$   &  $1.02\,10^{-19}\,\pm\,1.35\,10^{-20}$  &   $5$         &                     &                                          &               \\
    $2.80\,10^{4}$   &  $<\,1.07\,10^{-18}\,\,\,\,$            &   $5$         &                     &                                          &               \\
    $6.14\,10^{4}$   &  $1.91\,10^{-20}\,\pm\,6.35\,10^{-21}$  &   $5$         &                     &                                          &               \\
    $6.14\,10^{4}$   &  $1.27\,10^{-20}\,\pm\,2.44\,10^{-21}$  &   $5$         &                     &                                          &               \\
    $6.17\,10^{4}$   &  $<1.75\,10^{-20}\,\,\,\,$              &   $5$         &                     &                                          &               \\
\hline

\end{tabular}
\end{table*}

\begin{table*}
\caption{Photometric data used in this study for the targets
  \object{$\beta$~UMi} and \object{$\gamma$~Dra}. The literature
  references are specified at the end of Table~\ref{tablephotometry5}.}
\label{tablephotometry2}
\centering
\begin{minipage}{\textwidth}
 \centering
\begin{tabular}{|l|l|l||l|l|l|}
\hline
\multicolumn{3}{|c||}{$\beta$\,UMi} & \multicolumn{3}{c|}{$\gamma$\,Dra} \\
\hline\hline
 \multicolumn{1}{|c}{$\lambda$} & \multicolumn{1}{|c|}{$\lambda F_{\lambda}$} &  ref. & \multicolumn{1}{|c|}{$\lambda$} & \multicolumn{1}{c|}{$\lambda F_{\lambda}$} & ref. \\
 \multicolumn{1}{|c}{$[\mu\textrm{m}]$} & \multicolumn{1}{|c|}{$[\textrm{W} \textrm{m}^{-2}]$} & & \multicolumn{1}{|c|}{$[\mu\textrm{m}]$} & \multicolumn{1}{c|}{$[\textrm{W} \textrm{m}^{-2}]$} &  \\
\hline
    $3.46\,10^{-1}$  &  $8.12\,10^{-11}\,\pm\,8.73\,10^{-13}$   &   $4$        &        $3.46\,10^{-1}$  &  $5.58\,10^{-11}\,\pm\,4.45\,10^{-13}$   &   $4$        \\
    $3.60\,10^{-1}$  &  $1.17\,10^{-10}\,\pm\,0.00$             &   $1$        &        $3.60\,10^{-1}$  &  $8.84\,10^{-11}\,\pm\,0.00$             &   $1$        \\
    $4.01\,10^{-1}$  &  $3.91\,10^{-10}\,\pm\,4.19\,10^{-12}$   &   $4$        &        $4.01\,10^{-1}$  &  $2.99\,10^{-10}\,\pm\,2.38\,10^{-12}$   &   $4$        \\
    $4.23\,10^{-1}$  &  $7.79\,10^{-10}\,\pm\,7.17\,10^{-12}$   &   $4$        &        $4.23\,10^{-1}$  &  $6.17\,10^{-10}\,\pm\,4.02\,10^{-12}$   &   $4$        \\
    $4.40\,10^{-1}$  &  $1.15\,10^{-9}\,\pm\,0.00$              &   $1$        &        $4.40\,10^{-1}$  &  $9.65\,10^{-10}\,\pm\,0.00$             &   $1$        \\
    $4.48\,10^{-1}$  &  $1.24\,10^{-9}\,\pm\,1.33\,10^{-11}$    &   $4$        &        $4.48\,10^{-1}$  &  $1.01\,10^{-9}\,\pm\,8.04\,10^{-12}$    &   $4$        \\
    $5.39\,10^{-1}$  &  $2.78\,10^{-9}\,\pm\,2.99\,10^{-11}$    &   $4$        &        $5.39\,10^{-1}$  &  $2.35\,10^{-9}\,\pm\,1.88\,10^{-11}$    &   $4$        \\
    $5.49\,10^{-1}$  &  $3.01\,10^{-9}\,\pm\,2.22\,10^{-11}$    &   $4$        &        $5.49\,10^{-1}$  &  $2.57\,10^{-9}\,\pm\,1.18\,10^{-11}$    &   $4$        \\
    $5.50\,10^{-1}$  &  $3.06\,10^{-9}\,\pm\,0.00$              &   $1$        &        $5.50\,10^{-1}$  &  $2.69\,10^{-9}\,\pm\,0.00$              &   $1$        \\
    $5.81\,10^{-1}$  &  $3.66\,10^{-9}\,\pm\,3.93\,10^{-11}$    &   $4$        &        $5.81\,10^{-1}$  &  $3.19\,10^{-9}\,\pm\,2.54\,10^{-11}$    &   $4$        \\
    $7.00\,10^{-1}$  &  $5.28\,10^{-9}\,\pm\,0.00$              &   $1$        &        $7.00\,10^{-1}$  &  $4.77\,10^{-9}\,\pm\,0.00$              &   $1$        \\
    $9.00\,10^{-1}$  &  $6.68\,10^{-9}\,\pm\,0.00$              &   $1$        &        $9.00\,10^{-1}$  &  $6.56\,10^{-9}\,\pm\,0.00$              &   $1$        \\
    $1.24$           &  $6.10\,10^{-9}\,\pm\,1.09\,10^{-09}$    &   $12$       &        $1.24$           &  $4.84\,10^{-9}\,\pm\,9.46\,10^{-10}$    &   $12$       \\
    $1.25$           &  $6.43\,10^{-9}\,\pm\,0.00$              &   $6$        &        $1.25$           &  $6.37\,10^{-9}\,\pm\,0.00$              &   $6$        \\
    $1.25$           &  $5.64\,10^{-9}\,\pm\,2.01\,10^{-10}$    &   $7$        &        $1.25$           &  $5.46\,10^{-9}\,\pm\,2.20\,10^{-10}$    &   $7$        \\
    $1.66$           &  $5.82\,10^{-9}\,\pm\,9.86\,10^{-10}$    &   $12$       &        $1.66$           &  $4.88\,10^{-9}\,\pm\,8.09\,10^{-10}$    &   $12$       \\
    $2.16$           &  $3.03\,10^{-9}\,\pm\,5.68\,10^{-10}$    &   $12$       &        $2.16$           &  $2.70\,10^{-9}\,\pm\,3.97\,10^{-10}$    &   $12$       \\
    $2.20$           &  $3.09\,10^{-9}\,\pm\,0.00$              &   $6$        &        $2.20$           &  $2.97\,10^{-9}\,\pm\,0.00$              &   $6$        \\
    $2.20$           &  $2.85\,10^{-9}\,\pm\,3.32\,10^{-11}$    &   $7$        &        $2.20$           &  $2.80\,10^{-9}\,\pm\,4.24\,10^{-11}$    &   $7$        \\
    $3.50$           &  $9.18\,10^{-10}\,\pm\,2.94\,10^{-11}$   &   $7$        &        $3.40$           &  $1.09\,10^{-9}\,\pm\,0.00$              &   $1$        \\
    $3.50$           &  $8.69\,10^{-10}\,\pm\,2.94\,10^{-11}$   &   $7$        &        $3.40$           &  $1.03\,10^{-9}\,\pm\,0.00$              &   $6$        \\
    $4.20$           &  $6.02\,10^{-10}\,\pm\,1.66\,10^{-10}$   &   $2$        &        $3.50$           &  $9.18\,10^{-10}\,\pm\,2.45\,10^{-11}$   &   $7$        \\
    $4.90$           &  $2.92\,10^{-10}\,\pm\,6.45\,10^{-12}$   &   $7$        &        $3.50$           &  $8.69\,10^{-10}\,\pm\,2.45\,10^{-11}$   &   $7$        \\
    $4.90$           &  $2.82\,10^{-10}\,\pm\,6.45\,10^{-12}$   &   $7$        &        $4.20$           &  $6.60\,10^{-10}\,\pm\,1.82\,10^{-10}$   &   $2$        \\
    $1.10\,10^{1}$   &  $4.37\,10^{-11}\,\pm\,8.05\,10^{-12}$   &   $2$        &        $4.90$           &  $2.84\,10^{-10}\,\pm\,6.94\,10^{-12}$   &   $7$        \\
    $1.20\,10^{1}$   &  $2.82\,10^{-11}\,\pm\,5.65\,10^{-12}$   &   $3$        &        $4.90$           &  $2.74\,10^{-10}\,\pm\,6.94\,10^{-12}$   &   $7$        \\
    $1.98\,10^{1}$   &  $7.77\,10^{-12}\,\pm\,1.43\,10^{-12}$   &   $2$        &        $5.00$           &  $3.23\,10^{-10}\,\pm\,0.00$             &   $6$        \\
    $2.50\,10^{1}$   &  $3.26\,10^{-12}\,\pm\,6.52\,10^{-13}$   &   $3$        &        $1.02\,10^{1}$  &  $3.81\,10^{-11}\,\pm\,0.00$              &   $1$        \\
    $2.74\,10^{1}$   &  $4.17\,10^{-12}\,\pm\,1.15\,10^{-12}$   &   $2$        &        $1.02\,10^{1}$  &  $3.88\,10^{-11}\,\pm\,0.00$              &   $1$        \\
    $6.00\,10^{1}$   &  $2.17\,10^{-13}\,\pm\,4.35\,10^{-14}$   &   $3$        &        $1.10\,10^{1}$  &  $3.63\,10^{-11}\,\pm\,6.70\,10^{-12}$    &   $2$        \\
    $1.00\,10^{2}$   &  $5.73\,10^{-14}\,\pm\,1.15\,10^{-14}$   &   $3$        &        $1.20\,10^{1}$  &  $2.72\,10^{-11}\,\pm\,5.45\,10^{-12}$    &   $3$        \\
    $1.20\,10^{3}$   &  $4.00\,10^{-17}\,\pm\,1.00\,10^{-17}$   &   $5$        &        $1.98\,10^{1}$  &  $1.62\,10^{-11}\,\pm\,2.99\,10^{-12}$    &   $2$        \\
    $1.20\,10^{3}$   &  $3.05\,10^{-17}\,\pm\,6.00\,10^{-18}$   &   $10$       &        $2.50\,10^{1}$  &  $3.24\,10^{-12}\,\pm\,6.48\,10^{-13}$    &   $3$        \\
    $6.14\,10^{4}$   &  $5.37\,10^{-21}\,\pm\,0.00$             &   $5$        &        $6.00\,10^{1}$  &  $2.25\,10^{-13}\,\pm\,4.51\,10^{-14}$    &   $3$        \\
                     &                                          &              &        $1.00\,10^{2}$  &  $4.89\,10^{-14}\,\pm\,9.77\,10^{-15}$    &   $3$        \\
                     &                                          &              &        $3.50\,10^{2}$  &  $9.94\,10^{-16}\,\pm\,2.57\,10^{-16}$    &   $3$        \\
                     &                                          &              &        $1.20\,10^{3}$  &  $2.52\,10^{-17}\,\pm\,5.00\,10^{-18}$    &   $10$        \\
                     &                                          &              &        $1.20\,10^{3}$  &  $<\,3.00\,10^{-17}\,\,\,\,$              &   $5$        \\
                     &                                          &              &        $2.01\,10^{4}$  &  $<\,6.42\,10^{-20}\,\,\,\,$              &   $5$        \\
                     &                                          &              &        $3.55\,10^{4}$  &  $<\,1.27\,10^{-20}\,\,\,\,$              &   $5$        \\
\hline
\end{tabular}
\end{minipage}
\end{table*}

\begin{table*}
\caption{Photometric data used in this study for the targets
  \object{$\alpha$~Tau} and \object{$\beta$~And}. The literature
  references are specified at the end of Table~\ref{tablephotometry5}.}
\label{tablephotometry3}
\centering
\begin{minipage}{\textwidth}
 \centering
\begin{tabular}{|l|l|l||l|l|l|}
\hline
\multicolumn{3}{|c||}{$\alpha$\,Tau} & \multicolumn{3}{c|}{$\beta$\,And} \\
\hline\hline
 \multicolumn{1}{|c}{$\lambda$} & \multicolumn{1}{|c|}{$\lambda F_{\lambda}$} &  ref. & \multicolumn{1}{|c|}{$\lambda$} & \multicolumn{1}{c|}{$\lambda F_{\lambda}$} & ref. \\
 \multicolumn{1}{|c}{$[\mu\textrm{m}]$} & \multicolumn{1}{|c|}{$[\textrm{W} \textrm{m}^{-2}]$} & & \multicolumn{1}{|c|}{$[\mu\textrm{m}]$} & \multicolumn{1}{c|}{$[\textrm{W} \textrm{m}^{-2}]$} &  \\
\hline
$3.46\,10^{-1}$  &  $1.85\,10^{-10}\,\pm\,3.51\,10^{-12}$  &    $4$      &         $3.46\,10^{-1}$  &  $5.38\,10^{-11}\,\pm\,6.61\,10^{-13}$   &   $4$        \\
$3.60\,10^{-1}$  &  $2.93\,10^{-10}\,\pm\,0.00$            &    $1$      &         $3.60\,10^{-1}$  &  $9.17\,10^{-11}\,\pm\,0.00\,10^{+00}$   &   $1$        \\
$4.01\,10^{-1}$  &  $9.98\,10^{-10}\,\pm\,1.90\,10^{-11}$  &    $4$      &         $4.01\,10^{-1}$  &  $3.11\,10^{-10}\,\pm\,3.83\,10^{-12}$   &   $4$        \\
$4.23\,10^{-1}$  &  $2.09\,10^{-9}\,\pm\,3.47\,10^{-11}$   &    $4$      &         $4.23\,10^{-1}$  &  $6.60\,10^{-10}\,\pm\,5.98\,10^{-12}$   &   $4$        \\
$4.40\,10^{-1}$  &  $3.31\,10^{-9}\,\pm\,0.00$             &    $1$      &         $4.40\,10^{-1}$  &  $1.08\,10^{-9}\,\pm\,0.00$             &   $1$        \\
$4.48\,10^{-1}$  &  $3.37\,10^{-9}\,\pm\,6.40\,10^{-11}$   &    $4$      &         $4.48\,10^{-1}$  &  $1.07\,10^{-9}\,\pm\,1.31\,10^{-11}$    &   $4$        \\
$5.39\,10^{-1}$  &  $8.08\,10^{-9}\,\pm\,1.53\,10^{-10}$   &    $4$      &         $5.39\,10^{-1}$  &  $2.72\,10^{-9}\,\pm\,3.34\,10^{-11}$    &   $4$        \\
$5.49\,10^{-1}$  &  $8.79\,10^{-9}\,\pm\,1.21\,10^{-10}$   &    $4$      &         $5.49\,10^{-1}$  &  $2.94\,10^{-9}\,\pm\,1.08\,10^{-11}$    &   $4$        \\
$5.50\,10^{-1}$  &  $9.41\,10^{-9}\,\pm\,0.00$             &    $1$      &         $5.50\,10^{-1}$  &  $3.14\,10^{-9}\,\pm\,0.00$              &   $1$        \\
$5.81\,10^{-1}$  &  $1.07\,10^{-8}\,\pm\,2.04\,10^{-10}$   &    $4$      &         $5.81\,10^{-1}$  &  $3.59\,10^{-9}\,\pm\,4.41\,10^{-11}$    &   $4$        \\
$7.00\,10^{-1}$  &  $1.81\,10^{-8}\,\pm\,0.00$             &    $1$      &         $7.00\,10^{-1}$  &  $6.12\,10^{-9}\,\pm\,0.00$              &   $1$        \\
$9.00\,10^{-1}$  &  $2.71\,10^{-8}\,\pm\,0.00$             &    $1$      &         $9.00\,10^{-1}$  &  $9.65\,10^{-9}\,\pm\,0.00$              &   $1$        \\
$1.24$           &  $2.66\,10^{-8}\,\pm\,4.75\,10^{-9}$    &    $12$     &         $1.24$           &  $9.33\,10^{-9}\,\pm\,1.77\,10^{-09}$    &   $12$       \\
$1.25$           &  $2.31\,10^{-8}\,\pm\,0.00$             &    $1$      &         $1.25$           &  $8.68\,10^{-9}\,\pm\,2.92\,10^{-10}$    &   $7$        \\
$1.25$           &  $2.38\,10^{-8}\,\pm\,0.00$             &    $6$      &         $1.25$           &  $9.20\,10^{-9}\,\pm\,0.00$              &   $6$        \\
$1.66$           &  $2.43\,10^{-8}\,\pm\,3.80\,10^{-9}$    &    $12$     &         $1.66$           &  $8.80\,10^{-9}\,\pm\,1.20\,10^{-09}$    &   $12$       \\
$2.16$           &  $1.53\,10^{-8}\,\pm\,1.97\,10^{-9}$    &    $12$     &         $2.16$           &  $5.06\,10^{-9}\,\pm\,7.46\,10^{-10}$    &   $12$       \\
$2.20$           &  $1.14\,10^{-8}\,\pm\,0.00$             &    $1$      &         $2.20$           &  $4.83\,10^{-9}\,\pm\,5.64\,10^{-11}$    &   $7$        \\
$2.20$           &  $1.19\,10^{-8}\,\pm\,0.00$             &    $6$      &         $2.20$           &  $3.92\,10^{-9}\,\pm\,0.00$              &   $6$        \\
$3.40$           &  $4.33\,10^{-9}\,\pm\,0.00$             &    $6$      &         $3.40$           &  $1.87\,10^{-9}\,\pm\,0.00$              &   $1$        \\
$3.50$           &  $3.79\,10^{-9}\,\pm\,9.27\,10^{-11}$   &    $7$      &         $3.40$           &  $1.74\,10^{-9}\,\pm\,0.00$              &   $6$        \\
$3.50$           &  $3.65\,10^{-9}\,\pm\,9.27\,10^{-11}$   &    $7$      &         $3.50$           &  $1.56\,10^{-9}\,\pm\,3.78\,10^{-11}$    &   $7$        \\
$4.20$           &  $2.88\,10^{-9}\,\pm\,7.96\,10^{-10}$   &    $2$      &         $3.50$           &  $1.45\,10^{-9}\,\pm\,3.78\,10^{-11}$    &   $7$        \\
$4.90$           &  $1.19\,10^{-9}\,\pm\,1.30\,10^{-11}$   &    $7$      &         $4.20$           &  $9.54\,10^{-10}\,\pm\,2.64\,10^{-10}$   &   $2$        \\
$4.90$           &  $1.17\,10^{-9}\,\pm\,1.30\,10^{-11}$   &    $7$      &         $4.90$           &  $4.73\,10^{-10}\,\pm\,6.23\,10^{-12}$   &   $7$        \\
$5.00$           &  $1.24\,10^{-9}\,\pm\,0.00$             &    $1$      &         $4.90$           &  $4.65\,10^{-10}\,\pm\,6.23\,10^{-12}$   &   $7$        \\
$5.00$           &  $1.40\,10^{-9}\,\pm\,0.00$             &    $6$      &         $5.00$           &  $4.75\,10^{-10}\,\pm\,0.00$             &   $1$        \\
$1.02\,10^{1}$   &  $1.52\,10^{-10}\,\pm\,0.00$            &    $1$      &         $5.00$           &  $5.51\,10^{-10}\,\pm\,0.00$             &   $6$        \\
$1.02\,10^{1}$   &  $1.95\,10^{-10}\,\pm\,0.00$            &    $6$      &         $1.02\,10^{1}$   &  $6.94\,10^{-11}\,\pm\,0.00$             &   $1$        \\
$1.10\,10^{1}$   &  $1.74\,10^{-10}\,\pm\,4.81\,10^{-11}$  &    $2$      &         $1.02\,10^{1}$   &  $8.11\,10^{-11}\,\pm\,0.00$             &   $6$        \\
$1.20\,10^{1}$   &  $1.23\,10^{-10}\,\pm\,2.46\,10^{-11}$  &    $3$      &         $1.10\,10^{1}$   &  $7.59\,10^{-11}\,\pm\,1.40\,10^{-11}$   &   $2$        \\
$1.98\,10^{1}$   &  $3.09\,10^{-11}\,\pm\,5.70\,10^{-12}$  &    $2$      &         $1.20\,10^{1}$   &  $5.05\,10^{-11}\,\pm\,1.01\,10^{-11}$   &   $3$        \\
$2.50\,10^{1}$   &  $1.31\,10^{-11}\,\pm\,2.61\,10^{-12}$  &    $3$      &         $1.98\,10^{1}$   &  $1.12\,10^{-11}\,\pm\,2.07\,10^{-12}$   &   $2$        \\
$2.74\,10^{1}$   &  $9.55\,10^{-12}\,\pm\,2.64\,10^{-12}$  &    $8$      &         $2.50\,10^{1}$   &  $5.84\,10^{-12}\,\pm\,1.17\,10^{-12}$             &   $3$        \\
$6.00\,10^{1}$   &  $9.79\,10^{-13}\,\pm\,1.96\,10^{-13}$  &    $3$      &         $2.74\,10^{1}$   &  $4.57\,10^{-12}\,\pm\,1.26\,10^{-12}$   &   $2$        \\
$1.00\,10^{2}$   &  $1.76\,10^{-13}\,\pm\,3.52\,10^{-14}$  &    $3$      &         $6.00\,10^{1}$   &  $4.08\,10^{-13}\,\pm\,8.16\,10^{-14}$   &   $3$        \\
$3.50\,10^{2}$   &  $4.54\,10^{-15}\,\pm\,7.02\,10^{-16}$  &    $13$     &         $1.00\,10^{2}$   &  $8.39\,10^{-14}\,\pm\,1.68\,10^{-14}$   &   $3$        \\
$4.50\,10^{2}$   &  $2.03\,10^{-15}\,\pm\,4.00\,10^{-16}$  &    $13$     &         $1.20\,10^{3}$   &  $6.25\,10^{-17}\,\pm\,1.00\,10^{-17}$   &   $5$        \\
$1.20\,10^{3}$   &  $1.27\,10^{-16}\,\pm\,1.50\,10^{-17}$  &    $5$      &         $1.20\,10^{3}$   &  $<\,1.00\,10^{-16}\,\,\,\,$             &   $9$        \\
$1.38\,10^{3}$   &  $5.62\,10^{-17}\,\pm\,1.23\,10^{-17}$  &    $8$      &         $1.20\,10^{3}$   &  $5.87\,10^{-17}\,\pm\,1.10\,10^{-17}$   &   $10$       \\
$2.77\,10^{3}$   &  $1.51\,10^{-17}\,\pm\,1.58\,10^{-18}$  &    $8$      &         $1.33\,10^{4}$   &  $<\,2.02\,10^{-17}\,\,\,\,$             &   $5$        \\
$2.00\,10^{4}$   &  $8.98\,10^{-20}\,\pm\,1.50\,10^{-20}$  &    $5$      &         $2.01\,10^{4}$   &  $8.37\,10^{-20}\,\pm\,0.00$             &   $5$        \\
$2.01\,10^{4}$   &  $1.28\,10^{-19}\,\pm\,0.00$            &    $5$      &         $2.80\,10^{4}$   &  $<\,5.34\,10^{-19}\,\,\,\,$             &   $5$        \\
$2.80\,10^{4}$   &  $<\,5.34\,10^{-19}\,\,\,\,$            &    $5$      &         $3.55\,10^{4}$   &  $1.52\,10^{-20}\,\pm\,0.00$             &   $5$        \\
$3.55\,10^{4}$   &  $2.53\,10^{-20}\,\pm\,0.00$            &    $5$      &         $6.14\,10^{4}$   &  $<\,1.03\,10^{-20}\,\,\,\,$             &   $5$        \\
$6.14\,10^{4}$   &  $<\,1.32\,10^{-20}\,\,\,\,$            &    $5$      &        $6.97\,10^{5}$   &  $<\,4.30\,10^{-19}\,\,\,\,$             &   $5$        \\
$6.97\,10^{5}$   &  $<\,4.30\,10^{-19}\,\,\,\,$            &    $5$      &                         &            \\
\hline
\end{tabular}
\end{minipage}
\end{table*}

\begin{table*}
\caption{Photometric data used in this study for the targets
  \object{$\alpha$~Cet} and \object{$\beta$~Peg}. The literature
  references are specified at the end of Table~\ref{tablephotometry5}.}
\label{tablephotometry4}
\centering
\begin{minipage}{\textwidth}
 \centering
\begin{tabular}{|l|l|l||l|l|l|}
\hline
\multicolumn{3}{|c||}{$\alpha$\,Cet} & \multicolumn{3}{c|}{$\beta$\,Peg} \\
\hline\hline
 \multicolumn{1}{|c}{$\lambda$} & \multicolumn{1}{|c|}{$\lambda F_{\lambda}$} &  ref. & \multicolumn{1}{|c|}{$\lambda$} & \multicolumn{1}{c|}{$\lambda F_{\lambda}$} & ref. \\
 \multicolumn{1}{|c}{$[\mu\textrm{m}]$} & \multicolumn{1}{|c|}{$[\textrm{W} \textrm{m}^{-2}]$} & & \multicolumn{1}{|c|}{$[\mu\textrm{m}]$} & \multicolumn{1}{c|}{$[\textrm{W} \textrm{m}^{-2}]$} &  \\
\hline
 $3.46\,10^{-1}$  &  $3.44\,10^{-11}\,\pm\,5.67\,10^{-13}$  &    $4$    &     $3.46\,10^{-1}$  &  $3.36\,10^{-11}\,\pm\,1.80\,10^{-12}$   &   $4$          \\
 $3.60\,10^{-1}$  &  $5.68\,10^{-11}\,\pm\,0.00$            &    $1$     &    $3.60\,10^{-1}$  &  $5.95\,10^{-11}\,\pm\,0.00$             &   $6$          \\
 $4.01\,10^{-1}$  &  $1.98\,10^{-10}\,\pm\,3.27\,10^{-12}$  &    $4$     &       $3.60\,10^{-1}$  &  $6.23\,10^{-11}\,\pm\,0.00$             &   $1$          \\
 $4.23\,10^{-1}$  &  $4.16\,10^{-10}\,\pm\,5.70\,10^{-12}$  &    $4$     &       $4.01\,10^{-1}$  &  $2.16\,10^{-10}\,\pm\,1.16\,10^{-11}$   &   $4$          \\
 $4.40\,10^{-1}$  &  $6.49\,10^{-10}\,\pm\,0.00$            &    $1$     &       $4.23\,10^{-1}$  &  $4.42\,10^{-10}\,\pm\,2.32\,10^{-11}$   &   $4$          \\
 $4.48\,10^{-1}$  &  $6.82\,10^{-10}\,\pm\,1.13\,10^{-11}$  &    $4$     &       $4.40\,10^{-1}$  &  $6.99\,10^{-10}\,\pm\,0.00$             &   $6$          \\
 $5.39\,10^{-1}$  &  $1.82\,10^{-9}\,\pm\,3.00\,10^{-11}$   &    $4$     &       $4.40\,10^{-1}$  &  $6.61\,10^{-10}\,\pm\,0.00\,10^{+00}$   &   $1$          \\
 $5.49\,10^{-1}$  &  $1.98\,10^{-9}\,\pm\,2.01\,10^{-11}$   &    $4$     &       $4.48\,10^{-1}$  &  $7.17\,10^{-10}\,\pm\,3.84\,10^{-11}$   &   $4$          \\
 $5.50\,10^{-1}$  &  $2.02\,10^{-9}\,\pm\,0.00$             &    $1$     &       $5.39\,10^{-1}$  &  $2.01\,10^{-9}\,\pm\,1.07\,10^{-10}$    &   $4$          \\
 $5.81\,10^{-1}$  &  $2.39\,10^{-9}\,\pm\,3.95\,10^{-11}$   &    $4$     &       $5.49\,10^{-1}$  &  $2.15\,10^{-9}\,\pm\,1.11\,10^{-10}$    &   $4$          \\
 $7.00\,10^{-1}$  &  $4.35\,10^{-9}\,\pm\,0.00$             &    $1$     &       $5.50\,10^{-1}$  &  $2.24\,10^{-9}\,\pm\,0.00$              &   $6$          \\
 $9.00\,10^{-1}$  &  $7.95\,10^{-9}\,\pm\,0.00$             &    $1$     &       $5.50\,10^{-1}$  &  $2.08\,10^{-9}\,\pm\,0.00$              &   $1$          \\
 $1.24$           &  $7.51\,10^{-9}\,\pm\,9.90\,10^{-10}$   &    $12$    &       $5.81\,10^{-1}$  &  $2.53\,10^{-9}\,\pm\,1.36\,10^{-10}$    &   $4$          \\
 $1.25$           &  $7.11\,10^{-9}\,\pm\,0.00$             &    $1$     &       $7.00\,10^{-1}$  &  $5.53\,10^{-9}\,\pm\,0.00$              &   $6$          \\
 $1.25$           &  $7.56\,10^{-9}\,\pm\,2.40\,10^{-10}$   &    $7$     &       $7.00\,10^{-1}$  &  $5.48\,10^{-9}\,\pm\,0.00$              &   $1$          \\
 $1.66$           &  $8.27\,10^{-9}\,\pm\,1.39\,10^{-9}$    &    $12$    &       $9.00\,10^{-1}$  &  $1.17\,10^{-8}\,\pm\,0.00$              &   $6$          \\
 $2.20$           &  $3.85\,10^{-9}\,\pm\,0.00$             &    $1$     &       $9.00\,10^{-1}$  &  $1.15\,10^{-8}\,\pm\,0.00$              &   $1$          \\
 $2.20$           &  $4.10\,10^{-9}\,\pm\,4.63\,10^{-11}$   &    $7$     &       $1.24$           &  $1.45\,10^{-8}\,\pm\,2.58\,10^{-09}$    &   $12$         \\
 $3.40$           &  $1.36\,10^{-9}\,\pm\,0.00$             &    $1$     &       $1.25$           &  $1.20\,10^{-8}\,\pm\,4.31\,10^{-10}$    &   $7$          \\
 $3.40$           &  $1.53\,10^{-9}\,\pm\,0.00$             &    $1$     &       $1.25$           &  $1.10\,10^{-8}\,\pm\,4.31\,10^{-10}$    &   $7$          \\
 $3.50$           &  $1.33\,10^{-9}\,\pm\,4.36\,10^{-11}$   &    $7$     &       $1.66$           &  $1.34\,10^{-8}\,\pm\,2.22\,10^{-09}$    &   $12$         \\
 $3.50$           &  $1.26\,10^{-9}\,\pm\,4.36\,10^{-11}$   &    $7$     &       $2.16$           &  $8.27\,10^{-9}\,\pm\,1.17\,10^{-09}$    &   $12$         \\
 $4.20$           &  $9.54\,10^{-10}\,\pm\,1.76\,10^{-10}$  &    $2$     &       $2.20$           &  $6.51\,10^{-9}\,\pm\,0.00$              &   $1$          \\
 $4.90$           &  $4.08\,10^{-10}\,\pm\,9.49\,10^{-12}$  &    $7$     &       $2.20$           &  $6.98\,10^{-9}\,\pm\,9.16\,10^{-11}$    &   $7$          \\
 $4.90$           &  $3.86\,10^{-10}\,\pm\,9.49\,10^{-12}$  &    $7$     &       $3.40$           &  $2.42\,10^{-9}\,\pm\,0.00$              &   $1$          \\
 $5.00$           &  $3.64\,10^{-10}\,\pm\,0.00$            &    $1$     &       $3.40$           &  $2.56\,10^{-9}\,\pm\,0.00$              &   $1$          \\
 $5.00$           &  $4.58\,10^{-10}\,\pm\,0.00$            &    $6$     &       $3.50$           &  $2.25\,10^{-9}\,\pm\,5.76\,10^{-11}$    &   $7$          \\
 $1.02\,10^{1}$   &  $5.17\,10^{-11}\,\pm\,0.00$            &    $1$     &       $3.50$           &  $2.09\,10^{-9}\,\pm\,5.76\,10^{-11}$    &   $7$          \\
 $1.02\,10^{1}$   &  $6.10\,10^{-11}\,\pm\,0.00$            &    $6$     &       $4.20$           &  $1.51\,10^{-9}\,\pm\,4.18\,10^{-10}$    &   $2$          \\
 $1.10\,10^{1}$   &  $5.25\,10^{-11}\,\pm\,1.45\,10^{-11}$  &    $2$     &       $4.90$           &  $7.20\,10^{-10}\,\pm\,1.22\,10^{-11}$   &   $7$          \\
 $1.20\,10^{1}$   &  $4.12\,10^{-11}\,\pm\,8.24\,10^{-12}$  &    $3$     &       $4.90$           &  $6.69\,10^{-10}\,\pm\,1.22\,10^{-11}$   &   $7$          \\
 $1.98\,10^{1}$   &  $8.52\,10^{-12}\,\pm\,0.00$            &    $2$     &       $5.00$           &  $7.26\,10^{-10}\,\pm\,0.00$             &   $1$          \\
 $2.50\,10^{1}$   &  $4.78\,10^{-12}\,\pm\,9.57\,10^{-13}$  &    $3$     &       $5.00$           &  $8.26\,10^{-10}\,\pm\,0.00$             &   $6$          \\
 $6.00\,10^{1}$   &  $3.38\,10^{-13}\,\pm\,6.76\,10^{-14}$  &    $3$     &       $1.02\,10^{1}$  &  $9.94\,10^{-11}\,\pm\,0.00$              &   $1$          \\
 $1.00\,10^{2}$   &  $6.69\,10^{-14}\,\pm\,1.34\,10^{-14}$  &    $3$     &       $1.02\,10^{1}$  &  $1.24\,10^{-10}\,\pm\,0.00$              &   $6$          \\
 $3.50\,10^{2}$   &  $1.80\,10^{-15}\,\pm\,3.08\,10^{-16}$  &    $13$    &       $1.10\,10^{1}$  &  $1.00\,10^{-10}\,\pm\,2.77\,10^{-11}$    &   $2$          \\
 $4.50\,10^{2}$   &  $7.33\,10^{-16}\,\pm\,2.20\,10^{-16}$  &    $13$    &       $1.20\,10^{1}$  &  $6.82\,10^{-11}\,\pm\,1.36\,10^{-11}$    &   $3$          \\
 $1.20\,10^{3}$   &  $3.75\,10^{-17}\,\pm\,7.50\,10^{-18}$  &    $5$     &       $1.98\,10^{1}$  &  $1.95\,10^{-11}\,\pm\,0.00$              &   $2$          \\
 $1.20\,10^{3}$   &  $1.46\,10^{-16}\,\pm\,2.33\,10^{-17}$  &    $9$     &       $2.50\,10^{1}$  &  $8.29\,10^{-12}\,\pm\,1.66\,10^{-12}$    &   $3$          \\
 $2.01\,10^{4}$   &  $7.62\,10^{-20}\,\pm\,0.00$            &    $5$     &       $2.74\,10^{1}$  &  $6.61\,10^{-12}\,\pm\,0.00$              &   $2$          \\
 $3.55\,10^{4}$   &  $6.25\,10^{-20}\,\pm\,0.00$            &    $5$     &       $6.00\,10^{1}$  &  $5.95\,10^{-13}\,\pm\,1.19\,10^{-13}$    &   $3$          \\
 $3.56\,10^{4}$   &  $<\,1.43\,10^{-18}\,\,\,\,$            &    $5$     &       $1.00\,10^{2}$  &  $1.03\,10^{-13}\,\pm\,2.06\,10^{-14}$    &   $3$          \\
 $6.14\,10^{4}$   &  $<\,7.82\,10^{-21}\,\,\,\,$            &    $5$      &       $3.50\,10^{2}$  &  $3.09\,10^{-15}\,\pm\,3.17\,10^{-16}$    &   $13$          \\
 $1.31\,10^{5}$   &  $<\,1.01\,10^{-18}\,\,\,\,$            &    $5$     &       $4.50\,10^{2}$  &  $1.60\,10^{-15}\,\pm\,1.80\,10^{-16}$    &   $13$          \\
                  &                                         &            &       $1.20\,10^{3}$  &  $5.75\,10^{-17}\,\pm\,1.25\,10^{-17}$    &   $5$          \\
                  &                                         &            &       $1.20\,10^{3}$  &  $7.29\,10^{-17}\,\pm\,1.22\,10^{-17}$    &   $9$          \\
                  &                                         &            &       $1.20\,10^{3}$  &  $7.37\,10^{-17}\,\pm\,1.37\,10^{-17}$    &   $10$         \\
                  &                                         &            &       $7.00\,10^{3}$  &  $1.07\,10^{-18}\,\pm\,5.14\,10^{-20}$    &   $11$         \\
                  &                                         &            &       $2.01\,10^{4}$  &  $1.25\,10^{-19}\,\pm\,0.00$              &   $5$          \\
                  &                                         &            &       $3.55\,10^{4}$  &  $<\,3.04\,10^{-20}\,\,\,\,$              &   $5$          \\
                  &                                         &            &       $6.14\,10^{4}$  &  $<\,9.77\,10^{-21}\,\,\,\,$              &   $5$          \\

\hline
\end{tabular}
\end{minipage}
\end{table*}

\begin{table*}
  \caption{Photometric data used in this study for
    \object{$\sigma$~Lib}. The literature
    references are specified underneath the table.}
\label{tablephotometry5}
\centering
\begin{minipage}{\textwidth}
 \centering
\begin{tabular}{|l|l|l|}
\hline
\multicolumn{3}{|c|}{$\sigma$\,Lib} \\
\hline\hline
 \multicolumn{1}{|c}{$\lambda$} & \multicolumn{1}{|c|}{$\lambda F_{\lambda}$} &  ref. \\
 \multicolumn{1}{|c}{$[\mu\textrm{m}]$} & \multicolumn{1}{|c|}{$[\textrm{W} \textrm{m}^{-2}]$} & \\
\hline
    $3.46\,10^{-1}$  &  $1.62\,10^{-11}\,\pm\,5.96\,10^{-13}$   &   $4$        \\
    $3.60\,10^{-1}$  &  $2.82\,10^{-11}\,\pm\,0.00$             &   $6$        \\
    $3.60\,10^{-1}$  &  $2.62\,10^{-11}\,\pm\,0.00$             &   $1$        \\
    $4.01\,10^{-1}$  &  $9.69\,10^{-11}\,\pm\,3.57\,10^{-12}$   &   $4$        \\
    $4.23\,10^{-1}$  &  $1.98\,10^{-10}\,\pm\,6.70\,10^{-12}$   &   $4$        \\
    $4.40\,10^{-1}$  &  $3.08\,10^{-10}\,\pm\,0.00$             &   $1$        \\
    $4.40\,10^{-1}$  &  $3.28\,10^{-10}\,\pm\,0.00$             &   $6$        \\
    $4.48\,10^{-1}$  &  $3.24\,10^{-10}\,\pm\,1.19\,10^{-11}$   &   $4$        \\
    $5.39\,10^{-1}$  &  $9.09\,10^{-10}\,\pm\,3.35\,10^{-11}$   &   $4$        \\
    $5.49\,10^{-1}$  &  $9.77\,10^{-10}\,\pm\,2.97\,10^{-11}$   &   $4$        \\
    $5.50\,10^{-1}$  &  $1.02\,10^{-9}\,\pm\,0.00$              &   $1$        \\
    $5.50\,10^{-1}$  &  $1.08\,10^{-9}\,\pm\,0.00$              &   $6$        \\
    $5.81\,10^{-1}$  &  $1.16\,10^{-9}\,\pm\,4.26\,10^{-11}$    &   $4$        \\
    $7.00\,10^{-1}$  &  $2.60\,10^{-9}\,\pm\,0.00$              &   $1$        \\
    $7.00\,10^{-1}$  &  $2.96\,10^{-9}\,\pm\,0.00$              &   $6$        \\
    $9.00\,10^{-1}$  &  $5.35\,10^{-9}\,\pm\,0.00$              &   $1$        \\
    $9.00\,10^{-1}$  &  $5.98\,10^{-9}\,\pm\,0.00$              &   $6$        \\
    $1.24$           &  $5.19\,10^{-9}\,\pm\,7.84\,10^{-10}$    &   $12$       \\
    $1.25$           &  $5.65\,10^{-9}\,\pm\,0.00$              &   $6$        \\
    $1.25$           &  $5.20\,10^{-9}\,\pm\,1.40\,10^{-10}$    &   $7$        \\
    $1.66$           &  $5.70\,10^{-9}\,\pm\,9.13\,10^{-10}$    &   $12$       \\
    $2.16$           &  $3.35\,10^{-9}\,\pm\,6.12\,10^{-10}$    &   $12$       \\
    $2.20$           &  $3.20\,10^{-9}\,\pm\,0.00$              &   $6$        \\
    $2.20$           &  $3.06\,10^{-9}\,\pm\,4.83\,10^{-11}$    &   $7$        \\
    $3.40$           &  $1.18\,10^{-9}\,\pm\,0.00$              &   $1$        \\
    $3.50$           &  $1.03\,10^{-9}\,\pm\,2.60\,10^{-11}$    &   $7$        \\
    $3.50$           &  $9.74\,10^{-10}\,\pm\,2.60\,10^{-11}$   &   $7$        \\
    $4.20$           &  $6.02\,10^{-10}\,\pm\,1.11\,10^{-10}$   &   $2$        \\
    $4.90$           &  $3.23\,10^{-10}\,\pm\,7.76\,10^{-12}$   &   $7$        \\
    $4.90$           &  $3.11\,10^{-10}\,\pm\,7.76\,10^{-12}$   &   $7$        \\
    $5.00$           &  $3.23\,10^{-10}\,\pm\,0.00$             &   $1$        \\
    $5.00$           &  $3.60\,10^{-10}\,\pm\,0.00$             &   $6$        \\
    $1.02\,10^{1}$   &  $3.51\,10^{-11}\,\pm\,0.00$             &   $1$        \\
    $1.02\,10^{1}$   &  $4.42\,10^{-11}\,\pm\,0.00$             &   $6$        \\
    $1.10\,10^{1}$   &  $6.32\,10^{-11}\,\pm\,1.75\,10^{-11}$   &   $2$        \\
    $1.20\,10^{1}$   &  $3.55\,10^{-11}\,\pm\,7.10\,10^{-12}$   &   $3$        \\
    $1.98\,10^{1}$   &  $2.14\,10^{-11}\,\pm\,9.86\,10^{-12}$   &   $2$        \\
    $2.50\,10^{1}$   &  $3.69\,10^{-12}\,\pm\,7.39\,10^{-13}$   &   $3$        \\
    $6.00\,10^{1}$   &  $3.00\,10^{-13}\,\pm\,6.00\,10^{-14}$   &   $3$        \\
    $1.00\,10^{2}$   &  $6.84\,10^{-14}\,\pm\,1.37\,10^{-14}$   &   $3$        \\
    $1.20\,10^{3}$   &  $3.02\,10^{-17}\,\pm\,6.75\,10^{-18}$   &   $10$       \\
    $3.57\,10^{4}$   &  $<\,4.54\,10^{-19}\,\,\,\,$             &   $5$        \\
    $6.14\,10^{4}$   &  $<\,1.03\,10^{-20}\,\,\,\,$             &   $5$        \\
\hline
\end{tabular}
\end{minipage}
\flushleft
\footnotesize{
$^1$ UBVRIJKLMNH Photoelectric Catalogue \citep{morel1978},
$^2$ The Revised AFGL (RAFGL) Catalogue \citep{price1983},
$^3$ IRAS catalogue of Point Sources, Version 2.0,
$^4$ Observations in the Geneva Photometric System 4 \citep{rufener1989},
$^5$ Radio continuum emission from stars  \citep{wendker1995},
$^6$ Stellar Photometry in Johnson's 11-color system \citep{ducati2002},
$^7$ COBE DIRBE Point Source Catalog \citep{smith2004},
$^8$ \citet{cohen2005},
$^9$ \citet{dehaes2007}
$^{10}$ IRAM observations (this article)
$^{11}$ VLA observations (this article)
$^{12}$ 2MASS All-Sky Catalog of Point Sources \citep{skrutskie2006}
$^{13}$ CSO observations (this article)
}
\end{table*}

\end{appendix}

\end{document}